\documentclass{aa}  

\usepackage{graphicx}

\usepackage{txfonts}
\usepackage{hyperref}
\usepackage{float}
\newcommand{\radm}{\ {\rm rad \ m^{-2}}} 

\begin{document}

   \title{Using the polarization properties of double radio relics to probe the turbulent compression scenario}
   \titlerunning{The polarization of double radio relics}

   \author{C. Stuardi\inst{1,2},
          A. Bonafede\inst{1,2},
          K. Rajpurohit\inst{1,2},
          M. Br\"uggen\inst{3},
          F. de Gasperin\inst{2},
          D. Hoang\inst{3},
          R. J. van Weeren\inst{4}
          \and
          F. Vazza\inst{1,2,3}
          }
    \authorrunning{C. Stuardi et al.}

   \institute{Dipartimento di Fisica e Astronomia, Universit\`a di Bologna, via Gobetti 93/2, I-40129 Bologna, Italy\\
              \email{chiara.stuardi2@unibo.it}
         \and
             INAF - Istituto di Radioastronomia di Bologna, Via Gobetti 101, I-40129 Bologna, Italy
         \and
             Hamburger Sternwarte, Universit\"at Hamburg, Gojenbergsweg 112, 21029 Hamburg, Germany
         \and 
             Leiden Observatory, Leiden University, PO Box 9513, 2300 RA Leiden, The Netherlands
             }

   \date{Received XX; accepted YY}

 
  \abstract
   {Radio relics are Mpc-size synchrotron sources located in the outskirts of some merging galaxy clusters. Binary-merging systems with favorable orientation may host two almost symmetric relics, named double radio relics.}
   {Double radio relics are seen preferentially edge-on and, thus, constitute a privileged sample for statistical studies. Their polarization and Faraday rotation properties give direct access to the relics’ origin and magnetic fields.}
   {In this paper, we present a polarization and Rotation Measure (RM) synthesis study of four clusters hosting double radio relics, namely 8C 0212+703, Abell 3365, PLCK G287.0+32.9, previously missing polarization studies, and ZwCl 2341+0000, for which conflicting results have been reported. We used 1-2 GHz Karl G. Jansky Very Large Array observations. We also provide an updated compilation of known double radio relics with important observed quantities. We studied their polarization and Faraday rotation properties at 1.4 GHz and we searched for correlations between fractional polarization and physical resolution, distance from the cluster center, and shock Mach number.}
   {The weak correlations found between these quantities are well reproduced by state-of-the-art magneto-hydrodynamical simulations of radio relics, confirming that merger shock waves propagate in a turbulent medium with tangled magnetic fields. Both external and internal Faraday depolarization should play a fundamental role in determining the polarization properties of radio relics at 1.4 GHz. Although the number of double radio relics with RM information is still low, their Faraday rotation properties (i.e., rest-frame RM and RM dispersion below $40 \radm$ and non-Gaussian RM distribution) can be explained in the scenario in which shock waves with Mach numbers larger than 2.5 propagate along the plane of the sky and compress the turbulent intra-cluster medium.}
   {}


   \maketitle
%

\section{Introduction}

A large variety of diffuse synchrotron sources populates galaxy clusters. They unveil the non-thermal content of the intra-cluster medium (ICM): weak magnetic fields ($\sim 10-0.1 \mu$G) and relativistic particles. In particular, radio relics are observed in some galaxy clusters that have recently experienced a major merger as a consequence of hierarchical accretion processes \citep[see e.g.,][for a recent review]{vanWeeren19}.

Radio relics are Mpc-size synchrotron sources observed in the outskirts of a few galaxy clusters. They often show an arc-like shape, with the curvature pointing towards the cluster center, and high levels of fractional polarization (i.e., $> 20 \ \%$ at GHz frequencies). Their spectrum (defined by the flux density $S_\nu\propto\nu^{-\alpha}$) is steep, with $\alpha>1$, and often characterized by a steepening trend towards the cluster center. Double radio relics are a particular class of relics where two almost symmetric relics are observed on the opposite sides of the cluster center, along the main merger axis \citep[see e.g.,][]{Bonafede09a,deGasperin14,Bonafede17}.

It is well established that the origin of radio relics is connected with the presence of shocks injected in the ICM during the merger event \citep{Ensslin98}. Proof of this is the detection of surface brightness and/or temperature jumps in the X-ray observations of the majority of radio relics with suitable X-ray data \citep[e.g.,][]{Akamatsu13a} and, \textit{vice versa}, the detection of radio relics coincident with every X-ray detected cluster's shock \citep[see][for a recent detection]{Hlavacek18}. The emerging scenario is that shock waves are able to both accelerate the electrons responsible for the synchrotron emission, via Fermi I processes, and compress and amplify the magnetic field components along the shock plane \citep{Ensslin98,Hoeft07}. 

In this framework, it is expected that an idealized binary merger can generate two merger shock waves that travel into the opposite directions along the merger axis forming double radio relics \citep{Roettiger99,Ha18a}. A recent and comprehensive optical study confirmed that the merger axis of double relic galaxy clusters is preferentially near to the plane of the sky \citep{Golovich19a,Golovich19b}. Hence, double radio relics systems form an important sample because their merger geometry can be well constrained and projection effects on radio relics should be minimal since they are observed edge-on.

However, in this picture, a number of details are still missing. The major open question concerns the efficiency of the Diffusive Shock Acceleration (DSA) process which is invoked to accelerate particles from the thermal pool \citep{Jones91}. The predicted efficiency is not sufficient to produce the observed radio power considering the low Mach numbers ($M < 5$) measured from radio relics \citep[e.g.,][]{Botteon20a}. For this problem, there are two broad classes of solutions: one is the presence of mildly relativistic fossil electrons in the ICM, which provide the seeds for successive re-acceleration via DSA \citep{Pinzke13,Kang16,Inchingolo22}, the other involves processes of pre-acceleration of thermal electrons at the shock front \citep{Guo14a,Guo14b,Wittor20}. None of the two is actually validated to solve the efficiency problem. Other issues are the non-detection of $\gamma$-ray emission from galaxy clusters which would be also expected in case of DSA \citep{Vazza14,Vazza16}, and the radio spectral index of some relics that is incompatible with the DSA theory. The latter is the case of $\alpha<1$ \citep[see e.g., the southern relic in Abell 3667,][]{deGasperin22} and curved spectral index \citep[as observed in the fainter relics of the Toothbrush galaxy cluster,][]{Rajpurohit20a}.

Moreover, the role of magnetic fields in shaping radio relic emission is yet poorly understood. For example, it is questionable whether threads and filaments with an enhanced magnetic field strength could give origin to the filamentary structures observed in highly resolved images of radio relics \citep{Digennaro18,Rajpurohit22b,deGasperin22}. It is uncertain if magnetic fields can play an important role in particle acceleration since, for some mechanisms, the acceleration efficiency has a strong dependence on the pre-shock magnetic field alignment \citep[i.e.,][]{Guo14a,Caprioli14}. While it is known that intra-cluster magnetic fields can be amplified by a factor $\sim2$ by compression for $M\sim2-3$ shocks \citep{Iapichino12,Dominguez21a}, the numerous mechanisms that could lead to amplification in the low Mach number regime are still little explored from a theoretical point of view \citep{Donnert18}. Also, a quantitative estimate of magnetic field amplification at relics is difficult to obtain and the number of studies is limited \citep{Johnston-Hollitt04,Bonafede13,Stuardi21}.

A powerful tool to study magnetic fields in cluster radio relics is the analysis of their polarized emission. Since magnetic fields in relics are compressed and ordered along the shock plane, they are expected to be intrinsically highly polarized \citep{Ensslin98}. Their polarized emission carries fundamental information about their origin. In particular, their polarization properties (as the average fractional polarization and the spatial distribution of the fractional polarization across the relic) are strictly connected to the ICM turbulent properties and magnetic field structure \citep{Wittor19,Dominguez21b}. The direction of intrinsic polarization angle unveils the direction of the source magnetic field projected on the plane of the sky ($B_\perp$), while the rotation of the polarization angle with frequency, i.e. the Faraday rotation effect, depends on the magnetic field component of the magnetic field along the line-of-sight ($B_\parallel$) through the Rotation Measure (RM).

Following \citep{Burn66} we can express the polarization as a complex vector:

\begin{equation}
  \textbf{P} = Q+iU = P e^{i\chi} \ ,
\end{equation}

where $\chi$ is the polarization angle and $Q$ and $U$ are the Stokes parameters. The measured polarization angle depends on the observing wavelength squared, $\lambda^2$, and on the Faraday depth, $\phi$:

\begin{equation}
    \chi(\lambda^2) = \chi_0 + \phi\lambda^2 \ ,
\end{equation}

where $\chi_0$ is the intrinsic polarization angle of the radiation and the Faraday depth is defined as:

\begin{equation}
    \phi = 0.81 \int_{\rm source}^{\rm observer} {n_e B_\parallel} {\rm d}l \ {\rm [rad \ m^{-2}]}
    \label{eq:phi}
\end{equation}

with $n_e$, the thermal electron density, in cm$^{-3}$, $B_\parallel$ in $\mu$G and d$l$, the infinitesimal path length, in parsecs. The Rotation Measure is :

\begin{equation}
    RM = \frac{{\rm d}\chi(\lambda^2)}{{\rm d} \lambda^2} 
\end{equation}

and RM = $\phi$ only when $\chi$ and $\lambda^2$ are linearly correlated, i.e. when the Faraday rotation is caused by one or more (not emitting) screens in the source's foreground. This is often the case for radio relics, for which the measured RM is the sum of the Milky Way Faraday rotation and of the contribution from the external ICM. For this reason, RMs from relics can be used to define the relics position within the ICM and to infer the properties of the magnetic field in front of the relics themselves \citep{Pizzo11,Stuardi21,Rajpurohit22a}. When more complex Faraday depth structures are observed from radio relics, they are an indication of internal Faraday rotation and can be used to study the internal magneto-ionic structure of radio relics \citep{Stuardi19,Rajpurohit22a,deGasperin22}. Faraday effects may also cause wavelength-dependent depolarization \citep{Burn66}. Hence, the depolarization observed from relics is another important probe of magnetic field structure.

Polarization and, in particular, Faraday rotation studies of radio relics are still scarce in the literature. Only few bright radio relics have been studied in polarization with a good frequency coverage and physical resolution below 25 kpc \citep{Owen14,DiGennaro21,Rajpurohit22a,deGasperin22}. For most radio relics we only have information on their fractional polarization. This is true also for double radio relics, despite these systems may constitute a privileged sample because their geometry should favor the detection of their polarized emission \citep{Wittor19}.

Making a census of all double radio relics, we realized that three well-known radio relics totally miss radio polarization observations available in the literature, namely 8C 0212+703 (a.k.a. ClG 0217+70), Abell 3365 and PLCK G287.0+32.9 (a.k.a. PSZ2 G286.98+32.90). Hence, here we provide polarization and Faraday rotation images for these three galaxy clusters performed with $1-2$ GHz Karl G. Jansky Very Large Array (JVLA) observations. We decided also to analyze $1-2$ GHz JVLA observations of the double relic galaxy cluster ZwCl 2341.1+0000, for which polarization studies are already available but only at higher frequencies \citep{Benson17} or at low-resolution \citep{Giovannini10}. The main properties of the four clusters analyzed in this paper are listed in Tab.~\ref{tab:clusters}.

With this work, we want \textit{(i)} to increase the number of double radio relics with available polarization and Faraday rotation information and \textit{(ii)} to provide an insight into the polarization properties of all double radio relics known to date in order to probe their origin.

\begin{table}
    \centering
	\caption{Double relic galaxy clusters analyzed in this work. Column 1: name of the cluster; Column 2 and 3:  J2000 celestial coordinates retrieved from the NASA/IPAC Extragalactic Database (NED, \protect\url{https://ned.ipac.caltech.edu/}); Column 4: redshift, $z$, retrieved from NED with the exception of 8C 0212+703 for which an updated redshift is provided by \citet{Zhang20}; Column 5: Galactic Faraday rotation computed as the median of the Milky Way RM from \citet{Huts21} in a 1 degree diameter circle around the cluster position. The uncertainty is the average uncertainty within this circle.}
	\label{tab:clusters}
	\resizebox{\linewidth}{!}{
	\begin{tabular}{lcccc}
		\hline
		\hline
		\multicolumn{1}{c}{Cluster} & \multicolumn{1}{c}{R.A.} & \multicolumn{1}{c}{Dec} & \multicolumn{1}{c}{$z$} & Galactic RM \\
        \hline
        8C 0212+703 & 02$^\textrm{h}$17$^\textrm{m}$01$^\textrm{s}$ & +70$^{\circ}$36$\arcmin$.3 & 0.180 & $-24\pm21\radm$ \\
        Abell 3365 & 05$^{\rm h}$48$^{\rm m}$13$^{\rm s}$ & -21$^{\circ}$56$\arcmin$.1 & 0.093 & $26\pm10\radm$ \\
        PLCK G287.0+32.9 & 11$^{\rm h}$50$^{\rm m}$49$^{\rm s}$ & -28$^{\circ}$04$\arcmin$.6 & 0.390 & $-34\pm12\radm$ \\
        ZwCl 2341.1+0000 & 23$^{\rm h}$43$^{\rm m}$39$^{\rm s}$ & +00$^{\circ}$16$\arcmin$.7 & 0.270 & $-8\pm7\radm$ \\
		\hline
	\end{tabular}
	}
\end{table}

This paper is organized as follows: this introductory section is completed with a brief overview on available information for the four double relic galaxy clusters here analyzed; in Sec.~\ref{sec:obs} we present our radio observations and the polarization analysis; in Sec.~\ref{sec:result} we present our results; in Sec~\ref{sec:discuss} we discuss our results in comparison with magneto-hydrodynamical (MHD) simulations of radio relics and with an updated compilation of all double radio relics, while in Sec.~\ref{sec:conclusion} we summarize and draw the conclusion of our work. The broadband integrated radio spectra of a few double radio relics is computed in the Appendix~\ref{app:A}.

Throughout this paper, we assume a $\Lambda$CDM cosmological model, with $H_0$ = 69.6 km s$^{-1}$ Mpc$^{-1}$, $\Omega_\text{M}$ = 0.286, $\Omega_{\Lambda}$ = 0.714 \citep{Bennett14}.

\subsection{8C 0212+703 (ClG 0217+70)}
\label{sec:8C0212}

The radio diffuse emission of the galaxy cluster 8C 0212+703, hereafter 8C0212, was first discovered in the Westerbork Northern Sky Survey \citep[WENSS]{Rengelink97} by \citet{Delain06}. Comparing radio, X-ray, and optical data, \citet{Brown11} confirmed the presence of a central radio halo and of multiple radio relics. A recent study based on the spectroscopy of X-ray \textit{Chandra} data was able to revise the redshift of 8C0212 which is now established to be $z = 0.18$ \citep{Zhang20}. This made 8C0212 the galaxy cluster hosting the largest radio relic detected to date, with a projected linear size of 3.5 Mpc \citep{Hoang21}. 

The low-frequency radio emission of this cluster was studied by \citet{Hoang21} using the Low Frequency Array \citep[LOFAR]{vanHaarlem13}. Part of the data presented in this paper were also used by \citet{Hoang21} to make spectral index maps between 141 MHz and 1.5 GHz. This study confirmed the spectral index trend expected for relics both in the elongated western relic and in the spiral-like eastern one. \citet{Hoang21} found injection spectral indexes $\alpha_{\rm 141 MHz}^{\rm 1.5 GHz} = 0.72 \pm 0.05$ (for the western relic) and $\alpha_{\rm 141 MHz}^{\rm 1.5 GHz} = 1.14 \pm 0.07 , 0.93 \pm 0.08, 0.97 \pm 0.16 $ (for the three patches that compose the eastern relic) leading to shocks Mach number estimates ranging between 2.0 and 3.2. High-resolution radio images also found a possible connection between the emission of a radio galaxy and the diffuse radio emission nearby the eastern relic. No connection has been established between the radio halo emission and the X-ray detected discontinuities at the halo edges \citep{Zhang20}. 

\citet{Hoang21} did not provide polarization images of the diffuse radio sources. The detection of polarized emission would be a confirmation of the identification as radio relics . A detailed study of the X-ray emission at the position of the relics is also missing because the data used by \citet{Zhang19} only cover the central part of the cluster.

\subsection{Abell 3365}
\label{sec:A3365}

Abell 3365 \citep[$z$=0.093][]{Abell58,Struble99}, hereafter A3365, is a complex merging system little studied in the radio band. The eastern elongated radio relic was first discovered in the NRAO VLA Sky Survey \citep[NVSS,][]{Condon98} and then observed at 1.4 GHz with the Westerbork Synthesis Radio Telescope (WSRT) and VLA \citep{vanWeeren11c}. The latter study discovered a second radio relic in the north-west of the cluster whose identification was confirmed by the detection of an underlying shock front with $M = 3.9\pm0.8$ in the X-ray XMM-Newton images \citep{Urdampilleta21}. \citet{Urdampilleta21} discovered a second shock with $M = 3.5\pm0.6$ at the position of the eastern relic and a cold front at the western edge of the highly disturbed and NE-SW elongated cluster core. Optical galaxies are distributed in three main structures \citep{vanWeeren11c,Golovich19a,Golovich19b}: the most massive first component in the north-east has itself two merging sub-components which may have originated the eastern relic, the second western sub-component is going to merge with the third one that lies in the middle.
Recently, A3365 was observed with the Murchison Widefield Array (MWA) and the Australian Square Kilometre Array Pathfinder (ASKAP) by \citet{Duchesne21c} which were able to constrain the integrated spectral index of the eastern and western relics ($\alpha_\textrm{88 MHz}^\textrm{1.4 GHz} = 0.85\pm0.03$ and $\alpha_\textrm{118 MHz}^\textrm{1.4 GHz} = 0.76\pm0.08$, respectively). These estimates are incompatible with DSA theory.

\subsection{PLCK G287.0+32.9 (PSZ2 G286.98+32.90)}

PLCK G287.0+32.9, hereafter PLCK287, is an exceptionally luminous galaxy cluster at $z = 0.39$ detected by the Planck satellite \citep{Planck16b}. A pair of radio relics and a central radio halo were discovered by means of Giant Metrewave Radio Telescope (GMRT, at 150 MHz) and Very Large Array (1.4 GHz) observations by \citet{Bagchi11}. \citet{Bonafede14} performed a detailed multi-wavelength analysis of this cluster. New GMRT (at 325 and 610 MHz) and JVLA (2-4 GHz) radio images were used to study the radio spectral index of the two radio relics. Spectral index estimates were used to derive the Mach number of the two relics: $M \sim 3.7$ for the southern relic and $M \sim 5.4$ for the northern one. The northern relic revealed a connection with the emission of a radio galaxy and a peculiar spectral index profile that steepens along both the internal and external side of the relic. \citet{George17} also measured the integrated spectral index of the northern and southern relics obtaining $\alpha_{\rm 88 MHz}^{\rm 3 GHz} = 1.19 \pm 0.03$ and $\alpha_{\rm 88 MHz}^{\rm 3 GHz} = 1.36 \pm 0.04$, respectively.

PLCK287 is undergoing a major merger along the NW-SE direction, slightly misaligned with respect to the optically detected intergalactic filament where the cluster is located \citep{Bonafede14}. The different distances of the northern (400 kpc) and southern (2.8 Mpc) relic from the cluster center was used to infer a possible merging scenario where the southern relic was created by the first core passage towards the south while the northern relic originated in a second core-passage. Both the dynamical analysis of this cluster based on the optical spectroscopy \citep{Golovich19b} and the weak lensing analysis \citep{Finner17} found a weak signature of one (or multiple) sub-clusters nearby the southern radio relic. These components are not observed in the $10$ ks XMM-Newton observation presented in \citet{Bagchi11}. Overall, the dynamics of the merger is not clear, in particular concerning the origin of the southern bright radio relic.

\subsection{ZwCl 2341.1+0000}

ZwCl 2341.1+000, hereafter ZwCl2341, is the second most massive galaxy cluster of the Saraswati supercluster \citep{Bagchi17}. It is located at $z=0.270$ \citep{Golovich19b} along a filament of galaxies at $\sim$ 45 Mpc from the supercluster core. \citet{Bagchi02} first discovered the diffuse radio emission of this galaxy cluster using NVSS observations. They found that the radio emission likely originated from the formation process of a NW-SE elongated structure with a total extent of $\sim$ 6 Mpc that was also detected in the optical and X-ray observations. A detailed radio follow-up of this system was performed by \citet{vanWeeren09} using GMRT 610, 241, and 157 MHz images. They classified the northern and southern emissions as radio relics, although with a rather round shape. A tentative detection of a central extended emission connecting the radio relics was also reported at 1.4 GHz, first by \citet{Giovannini10}, using VLA observations, and more recently by \citet{Parekh22} with the MeerKAT radio telescope \citep{Jonas16}.

\citet{Giovannini10} also reported polarized emission from the whole region of extended radio emission but, due to the very low-resolution ($83\arcsec\times75\arcsec$), the emission of the relics could be blended with other cluster sources and subject to beam depolarization. They obtained a $15 \ \%$ average polarization fraction for the northern relic and $8 \ \%$ for the southern one at 1.4 GHz. \citet{Benson17} published polarization images of ZwCl2341 using JVLA 2-4 GHz observations and obtained much lower average polarization fractions: $5 \ \%$ for the northern relic and $8 \ \%$ for the southern one which also shows a maximum polarization fraction of $30 \ \%$. Since higher fractional polarization is expected at higher frequencies, due to wavelength-dependent depolarization effects, a polarization study at 1.4 GHz at higher resolution is needed in order to investigate the discrepancy between these two results.

Several optical studies of this system \citep{Boschin13,Benson17,Golovich19b} found that it is composed of at least three sub-clusters: two of them are aligned along the NW-SE elongation of the X-ray emission and their merger is possibly responsible for the radio relics formation, while the third one in the north-east is likely to be involved in a secondary merger along the line-of-sight. \citet{Zhang21} performed a detailed analysis of a deep 206.5 ks \textit{Chandra} observation of ZwCl2341. They discovered the presence of numerous substructures within this cluster and confirmed its complex dynamical state. They could not detect shocks underlying the radio relics \citep[as previously attempted by][]{Akamatsu13a,Ogrean14} but they found a surface brightness edge at the position of the southern relic, which they interpreted as a kink due to the disrupted core of the southern sub-cluster. The northern relic lies instead at the apex of a conic X-ray structure delimited by cold fronts on both sides. \citet{Zhang21} also presented resolved spectral index maps between 325 MHz GMRT and 1.5 GHz JVLA observations (the same that are used in this work). Both relics show a spectral steeping towards the center of the cluster although the trend is not very clear, also due to the patchy shape of the two relics. From the injection spectral index they estimated a radio Mach number $M= 2.2 \pm 0.1$ and $M = 2.4 \pm 0.4$ for the southern and northern relic, respectively.

\begin{table*}
	\centering
	\caption{Details of the observations. Column 1: name of the cluster with a subscript letter that specifies the relic whenever necessary; Column 2 and 3:  J2000 celestial coordinates pointed in the observation; Column 4: array configuration; Column 5: observing date; Column 6: total integration time on each source.}
	\label{tab:observations}
	\begin{tabular}{lccccc}
		\hline
		\hline
		\multicolumn{1}{c}{Cluster$_{\rm relic}$} & \multicolumn{1}{c}{R.A.} & \multicolumn{1}{c}{Dec} & \multicolumn{1}{c}{Array Conf.} & \multicolumn{1}{c}{Obs. Date} & \multicolumn{1}{c}{Time on source [h]} \\
        \hline
        8C 0212+703$_E$ & 02$^{\rm h}$18$^{\rm m}$50.0$^{\rm s}$ & +70$^{\circ}$27$\arcmin$36.0$\arcsec$ & C & 03 Jun 2017 & 1.7 \\
        8C 0212+703$_E$ & 02$^{\rm h}$18$^{\rm m}$50.0$^{\rm s}$ & +70$^{\circ}$27$\arcmin$36.0$\arcsec$ & D & 21 Mar 2017 & 0.8 \\
        8C 0212+703$_W$ & 02$^{\rm h}$14$^{\rm m}$31.0$^{\rm s}$ & +70$^{\circ}$41$\arcmin$04.0$\arcsec$ & C & 03 Jun 2017 & 1.7 \\
        8C 0212+703$_W$ & 02$^{\rm h}$14$^{\rm m}$31.0$^{\rm s}$ & +70$^{\circ}$41$\arcmin$04.0$\arcsec$ & D & 21 Mar 2017 & 0.8 \\
        \hline
        Abell 3365$_E$ & 05$^{\rm h}$49$^{\rm m}$04.0$^{\rm s}$ & -21$^{\circ}$47$\arcmin$05.0$\arcsec$ & C & 29 May 2017 & 1.7 \\
        Abell 3365$_E$ & 05$^{\rm h}$49$^{\rm m}$04.0$^{\rm s}$ & -21$^{\circ}$47$\arcmin$05.0$\arcsec$ & D & 21 Mar 2017 & 0.4 \\
        Abell 3365$_W$ & 05$^{\rm h}$48$^{\rm m}$04.0$^{\rm s}$ & -21$^{\circ}$52$\arcmin$40.0$\arcsec$ & C & 29 May 2017 & 1.7 \\
        Abell 3365$_W$ & 05$^{\rm h}$48$^{\rm m}$04.0$^{\rm s}$ & -21$^{\circ}$52$\arcmin$40.0$\arcsec$ & D & 21 Mar 2017 & 0.4 \\
        \hline
        PLCK G287.0+32.9 & 11$^{\rm h}$51$^{\rm m}$00.0$^{\rm s}$ & -28$^{\circ}$07$\arcmin$17.0$\arcsec$ & C & 03 Jun 2017 & 1.7 \\
        PLCK G287.0+32.9 & 11$^{\rm h}$51$^{\rm m}$00.0$^{\rm s}$ & -28$^{\circ}$07$\arcmin$17.0$\arcsec$ & D & 15 Feb 2017 & 0.3 \\
        \hline
        ZwCl 2341.1+0000 & 23$^{\rm h}$43$^{\rm m}$39.7$^{\rm s}$ & +00$^{\circ}$16$\arcmin$39.0$\arcsec$ & C & 02 Jul 2017 & 1.6 \\
        ZwCl 2341.1+0000 & 23$^{\rm h}$43$^{\rm m}$39.7$^{\rm s}$ & +00$^{\circ}$16$\arcmin$39.0$\arcsec$ & D & 18 Feb 2017 & 0.8 \\
        ZwCl 2341.1+0000 & 23$^{\rm h}$43$^{\rm m}$44.0$^{\rm s}$ & +00$^{\circ}$17$\arcmin$18.0$\arcsec$ & C & 31 Jan 2016 & 3.2 \\
		ZwCl 2341.1+0000 & 23$^{\rm h}$43$^{\rm m}$44.0$^{\rm s}$ & +00$^{\circ}$17$\arcmin$18.0$\arcsec$ & D & 16 Oct 2015 & 1.4 \\
		\hline
	\end{tabular}
\end{table*}

\section{Radio observations}
\label{sec:obs}

The four clusters have been observed with the JVLA in the L-band ($1-2$ GHz) within the observing proposal 17A-083. In the case of ZwCl2341, we also analyzed two archival observations collected under the observing proposal SG0365. We used C- and D-configuration observations. 8C0212 and Abell 3365 were observed with two separated pointings on the two relics to maximize the sensitivity in the region of interest. The pointing center of each observation, the array configuration, and the observing date and time are summarized in Tab.~\ref{tab:observations}. The L-band spans 1024 MHz, covered by 16 spectral windows of 64 MHz (and 64 1-MHz-channels) each. Full polarization products have been recorded.

\subsection{Data reduction}
\label{sec:cal}

The dataset were automatically pre-processed right after the observation with the VLA \texttt{CASA}\footnote{\url{https://casa.nrao.edu/}} calibration pipeline (version \texttt{4.7.1} for D-array and \texttt{4.7.2} for C-array observations). This pipeline is optimized for Stokes I continuum data and it performs standard flagging and calibration procedures. We used the \texttt{CASA 5.6.1} package to complete the calibration also for the cross-correlation polarization products and to perform additional flagging.

We used the \citet{Perley13} flux density scale for wide-band observations as a model for the primary calibrator of each observation. To build a frequency-dependent polarization model we made a polynomial fit to the values of linear polarization fraction and polarization angle of a polarized calibrator, following the NRAO polarimetry guide for polarization calibration\footnote{\url{https://science.nrao.edu/facilities/vla/docs/manuals/obsguide/modes/pol}}. An unpolarized source was used to calibrate the on-axis instrumental leakage. The final calibration tables were applied to the target. 

Radio frequency interference (RFI) was removed manually and using statistical flagging algorithms also from the cross-correlation products. Some spectral windows were entirely removed. In particular spectral windows 1, 2, 3, 8, and 9 were often severely affected by RFI. The calibrated data were then averaged in time down to 10 s and in frequency with channels of 4 MHz, in order to speed up the subsequent imaging and self-calibration processes. We computed new visibility weights according to the visibilities scatter.

We used the multi-scale multi-frequency de-convolution algorithm of the \texttt{CASA} task \texttt{tclean} \citep{Rau11} for wide-band synthesis-imaging. As a first step, we made a large image of the entire field of view ($\sim1^{\circ}\times1^{\circ}$). We used a three Taylor expansion (\texttt{nterms} = 3) in order to take into account both the source spectral index and the primary beam response at large distances from the pointing center. For C-configuration data we used the $w$-projection algorithm \citep{Cornwell08} to correct for the wide-field non-coplanar baseline effect using 128 $w$-projection planes. The large images were recursively improved performing several cycles of self-calibration. This is the standard process to refine the antenna-based phase gain variations. During the last cycle, amplitude gains were also computed and applied, if possible.

In order to reduce the noise generated by bright sources in the field and to speed up the subsequent imaging processes, we subtracted all the sources external to the field of interest ($\sim15\arcmin\times15\arcmin$) from the visibilities. Since the subtraction is not applied to cross-correlation products, polarized sources will be present outside the field of interest. This is not a problem since, both, the polarized flux density and the number of polarized sources are lower with respect to the total intensity. After the subtraction, we reduced the number of $w$-projection planes to 64, and we used a Briggs weighting scheme with the \texttt{robust} parameter set to 0.5. The latter choice was done to better image the extended emission. In the case of 8C0212, we also subtracted a bright source at the J2000 sky coordinates [02$^{\rm h}$14$^{\rm m}$32.3$^{\rm s}$; +70$^{\circ}$49$\arcmin$16.7$\arcsec$], because its variability between the the times of C- and D-configuration observations caused imaging artifacts.

We performed a final cycle of phase and amplitude self-calibration using together the C- and D- configuration. Only the C-configuration observation was used for the analysis of PLCK287 because the addition of the D-configuration resulted in a loss of resolution and did not improve the final image quality. 

In the case of ZwCl2341, we also combined the two archival observations made in the previous observing cycle. The pointing center of these observations is $1.3\arcmin$ offset with respect to our observations. We checked that the flux density difference between the two observations due to the primary beam response is $\sim0.5\%$. Since this difference is well below our residual calibration errors on the amplitude ($\sim5 \ \%$), we simply shifted the phase center to the one of our observations. 

The primary beam image was obtained with the \texttt{widebandpbcor} task in \texttt{CASA} and then used to correct the final images.

\subsection{Polarization and RM synthesis}
\label{sec:im}

To produce final images of the Stokes parameters ($I$, $Q$ and $U$) for the polarization analysis, we used \texttt{WSCLEAN 3.0.1}\footnote{\url{https://gitlab.com/aroffringa/wsclean}} \citep{Offringa14,Offringa17}. This package is optimized to produce the wide-field frequency cubes, that will be used for the RM synthesis, as well as to take care of the wide-band to produce the images integrated over the full-band.

We produced image cubes with 64 channels at a frequency resolution of 16 MHz each.  We also produced Stokes $I$, $Q$ and $U$ images integrated over the full-band. The Stokes $Q$ and $U$ images were cleaned together using the \texttt{join-channels} and \texttt{join-polarizations} options. The large-scale emission of radio relics was modeled using the automated \texttt{multi-scale} option. We notice that the \texttt{multi-scale} algorithm, which is necessary to optimally image large scale emission, is not well implemented to work with the  \texttt{squared-channel-joining} algorithm that should be the preferable option for cleaning Stokes $Q$ and $U$. We used the Briggs weighting scheme with \texttt{robust} = 0.5. In order to perform the RM synthesis the restoring beam was forced to be the same in the full-band image and in each frequency channels, matching the lowest resolution one (i.e., at 1.02 GHz). This is done to avoid frequency-depended effects due to the variable beam size. However, we also created Stokes $I$ images at full-resolution and in Tab.~\ref{tab:pol} we listed the characteristics of both full-resolution and low-resolution total intensity images. Some frequency channels were discarded due to their higher noise with respect to average rms noise in the other channels. Finally, each image was corrected for the primary beam calculated with \texttt{CASA} for the central frequency of each channel. The details of the images created in this section are listed in Tab.~\ref{tab:pol}.

\begin{table*}
    \centering
	\caption{Details of the images. Column 1: name of the cluster with a subscript letter that specifies the relic whenever necessary; Column 2: JVLA array configuration used to produce the image; Column 3: robust parameter used for the Briggs weighting scheme; Column 4: FWHM of the tapering function used to produce the image; Column 5: FWHM of the major and minor axis of the resolution beam of the image; Column 6: rms noise of the total intensity image; Column 7: rms noise of the polarized intensity image produced from the Q and U images integrated over the full-band; Column 8: average noise in the polarized intensity image computed through RM-synthesis. See Sec.~\ref{sec:im} for the details on the noise calculation. Column 9: reference to the figure in this paper.}
	\label{tab:pol}
	\begin{tabular}{lcccccccc} 
		\hline
		\hline
		Cluster$_{\rm relic}$ &Array Conf. & Robust & Taper & Beam & $\sigma_I$ & $\sigma_{QU}$ & <$\sigma_{\widetilde{Q}\widetilde{U}}$> & Fig. \\
		       &            &  &  &  & [$\mu$Jy/beam] & [$\mu$Jy/beam] & [$\mu$Jy/beam/RMSF] & \\
		\hline

		8C 0212+703$_E$ & C+D  & 0.5 & -- & 18.2$\arcsec\times$12.2$\arcsec$ & 20 & -- & -- & \ref{fig:8C0212_nice},\ref{fig:8C0212_polE} \\
		    & C+D    & 0.5 & 20$\arcsec$ &  29$\arcsec\times$29$\arcsec$ & 35 & 18 & 15 & \ref{fig:8C0212_nice},\ref{fig:8C0212_polE} \\
		8C 0212+703$_W$ & C+D  & 0.5 & -- & 18.6$\arcsec\times$12.8$\arcsec$ & 22 & -- & -- & \ref{fig:8C0212_nice},\ref{fig:8C0212_polW} \\
		    & C+D    & 0.5 & 20$\arcsec$ &  33$\arcsec\times$33$\arcsec$ & 35 & 15 & 17 & \ref{fig:8C0212_nice},\ref{fig:8C0212_polW} \\
		Abell 3365$_E$ & C+D & 0.5 & -- &  21.0$\arcsec\times$10.6$\arcsec$ & 28 & -- & -- & \ref{fig:A3365_nice}, \ref{fig:A3365_pol}  \\
		    & C+D    & 0.5 & -- &  30$\arcsec\times$30$\arcsec$ & 28 & 13 & 16 & \ref{fig:A3365_nice}, \ref{fig:A3365_pol}  \\
		Abell 3365$_W$ & C+D  & 0.5 & -- & 21.0$\arcsec\times$10.7$\arcsec$ & 25 & -- & -- & \ref{fig:A3365_nice} \\
		    & C+D    & 0.5 & -- &  31$\arcsec\times$31$\arcsec$ & 28 & 16 & 19 & \ref{fig:A3365_nice} \\
		PLCK G287.0+32.9 & C & 0.5 & -- &  28.4$\arcsec\times$11.7$\arcsec$ & 33 & -- & -- & \ref{fig:PLCK287_nice}, \ref{fig:PLCK287_pol} \\
		    &  C    & 0.5 & -- &  41$\arcsec\times$41$\arcsec$ & 50 & 11 & 14 & \ref{fig:PLCK287_nice}, \ref{fig:PLCK287_pol} \\
		ZwCl 2341.1+0000 & C+D & 0.5 & -- & 16.3$\arcsec\times$13.5$\arcsec$ & 17 & -- & -- & \ref{fig:ZwCl2341_nice}, \ref{fig:ZwCl2341_pol_N}, \ref{fig:ZwCl2341_pol_S} \\
		    & C+D    & 0.5 & -- &  28$\arcsec\times$28$\arcsec$ & 18 & 8 & 11 & \ref{fig:ZwCl2341_nice}, \ref{fig:ZwCl2341_pol_N}, \ref{fig:ZwCl2341_pol_S} \\
		\hline
	\end{tabular}
\end{table*}

We refer to \citet{Brentjens05} for a comprehensive introduction to the RM-synthesis technique. In practice, the RM synthesis performs a Fourier transform of the wavelength-squared-dependent polarization into Faraday space, obtaining the polarization as a function of the Faraday depth, $\phi$ (see Eq.~\ref{eq:phi}). In Faraday space the polarization has a peak at the Faraday depth that rotates the polarization angle of the emission. In the following, we will refer to $\phi$ to describe the Faraday space in which the RM synthesis is performed, but we will use the more common term RM to describe the actual value derived applying this technique. This is possible because we did not detect Faraday-complex sources, for which RM and $\phi$ do not coincide.

Similarly to the observing beam of an interferometric image, the Rotation Measure Sampling Function (RMSF) represents the instrumental response to a polarized signal in Faraday space. While the observing beam depends on the antenna configuration, the RMSF depends on the observational bandwidth and on the width of the sub-bands in the $\lambda^2$-space. \citet{Brentjens05} obtained approximated formulas to compute the resolution in Faraday space, $\delta\phi$, the maximum observable Faraday depth, $|\phi_\text{max}|$, and the largest observable scale in Faraday space, $\Delta\phi_\text{max}$ (i.e., the depth and the $\phi$-scale at which sensitivity has dropped to $50 \ \%$). Therefore, considering our observing parameters, we have:

\begin{align}
    &\delta\phi \sim 45 \ \rm rad\,m^{-2} \ ,\\
    &|\phi_\text{max}| \sim 535 \ \rm rad\,m^{-2} \ ,\\
    &\Delta\phi_\text{max} \sim 143 \ \rm rad\,m^{-2} \ .
\end{align}

We performed the RM synthesis on the $Q(\nu)$ and $U(\nu)$ frequency cubes with \texttt{pyrmsynth}\footnote{\url{https://github.com/mrbell/pyrmsynth}}. Faraday cubes were created between $\pm1000\radm$ in order to have a wide range outside our $|\phi_\text{max}|$ to compute the noise from the spectra. We used equal weights for all the channels and we imposed a spectral correction using an average spectral index $\alpha$ = 1. We masked the $Q(\nu)$ and $U(\nu)$ images using the $3\sigma$ threshold applied to the full-band total intensity images having the same resolution as the frequency cubes (see Tab.~\ref{tab:pol} for the rms noise of these images). We also performed the RM clean down to the same threshold \citep[see][for the RM clean technique]{Heald09}. 

Applying the RM synthesis we obtained, in each pixel of the image, the reconstructed Faraday dispersion function, or Faraday spectrum, $\widetilde{F}(\phi)$, which describes the polarization as a function of the Faraday depth. We also obtained the reconstructed $\widetilde{Q}(\phi)$, $\widetilde{U}(\phi)$ cubes in Faraday space. For each pixel we measured the noise of $\widetilde{Q}(\phi)$ and $\widetilde{U}(\phi)$ computing the rms, $\sigma_{\widetilde{Q}}$ and $\sigma_{\widetilde{U}}$, in the external ranges of the spectrum: at $|\phi| \ > \ 500\radm$. This Faraday depth range is outside of the sensitivity range of our observations and should be free from the contamination of residual side-lobes. Since $\sigma_{\widetilde{Q}}\sim\sigma_{\widetilde{U}}$, we estimated the noise of each pixel of the polarization observations as $\sigma_{\widetilde{Q}\widetilde{U}}=(\sigma_{\widetilde{Q}}+\sigma_{\widetilde{U}})/2$ \citep[see also][]{Hales12}. By definition, $\sigma_{\widetilde{Q}\widetilde{U}}$ is in units of Jy/beam/RMSF. In Tab.~\ref{tab:pol}, we list the value of <$\sigma_{\widetilde{Q}\widetilde{U}}$>, where the average is computed over the image using all the unmasked pixels.

We fitted pixel-by-pixel a parabola around the main peak of the Faraday dispersion function. From the value of the Faraday depth at the peak we obtained the RM = $\phi_{\rm peak}$, while from $|\widetilde{F}(\phi_{\rm peak})|$ we obtained the polarized intensity. For our analysis, we considered only pixels with $|\widetilde{F}(\phi_{\rm peak})|$ > 6$\sigma_{\widetilde{Q}\widetilde{U}}$. This corresponds to a Gaussian significance level of about 5$\sigma$ \citep[see][]{Hales12}.

We computed polarization intensity images using the peak of the Faraday dispersion function, and correcting for the Ricean bias as $P=\sqrt{|\widetilde{F}(\phi_{\rm peak})|^2-2.3\sigma_{\widetilde{Q}\widetilde{U}}^2}$ \citep{George12}. We then obtained fractional polarization images dividing the $P$ images (with the 6$\sigma_{\widetilde{Q}\widetilde{U}}$ threshold) by the full-band Stokes $I$ images (masked at the 3$\sigma$ level). In order to check the RM synthesis results we also produced integrated polarization images using the $Q$ and $U$ images obtained using the full-band and applying the usual formula $P = \sqrt{Q^2 + U^2}$. The rms noise of these images is listed in Tab.~\ref{tab:pol} as $\sigma_{QU}$.

In order to compute the upper limits to the fractional polarization for those sources which were not detected in polarization we created a map of 6$\sigma_{\widetilde{Q}\widetilde{U}}$/$I$. We considered as an upper limit the minimum of all values reached in this map within the relic region. This upper limit is more stringent than the one calculated from the average value across the source, but it is comparable with fractional polarization values which are computed only from the brightest pixels.

From the reconstructed values of $Q$ and $U$ at $\phi_{\rm peak}$ we can also recover the intrinsic polarization angle (i.e., corrected for the value of RM determined by $\phi_{\rm peak}$), $\chi_0$, as:
\begin{equation}
 \label{eq:chi0}
     \chi_0=\chi(\lambda_0^2)-\phi_{\rm peak}\lambda_0^2=\frac{1}{2}\arctan{\frac{\widetilde{U}(\phi_{\rm peak})}{\widetilde{Q}(\phi_{\rm peak})}}-\phi_{\rm peak}\lambda_0^2~,
\end{equation}
where $\lambda_0$ is the central wavelength in the sampled wavelength-squared space. The magnetic field projected on the plane of the sky is then obtained from $\chi_0$.

The pixel-wise uncertainty on $\phi_{\rm peak}$ (and thus on the RM value in the single pixel) is derived following \citet{Brentjens05}, where:

\begin{equation}
    \sigma_{\phi}=\frac{\delta\phi}{2 P/\sigma_{\widetilde{Q}\widetilde{U}}}~,
\label{eq:errphi}
\end{equation}
 
that is the FWHM of the RMSF divided by twice the signal-to-noise of the detection \citep[see also][]{Schnitzeler17}.

\begin{figure*}[h]
    \centering
	\includegraphics[width=0.9\textwidth]{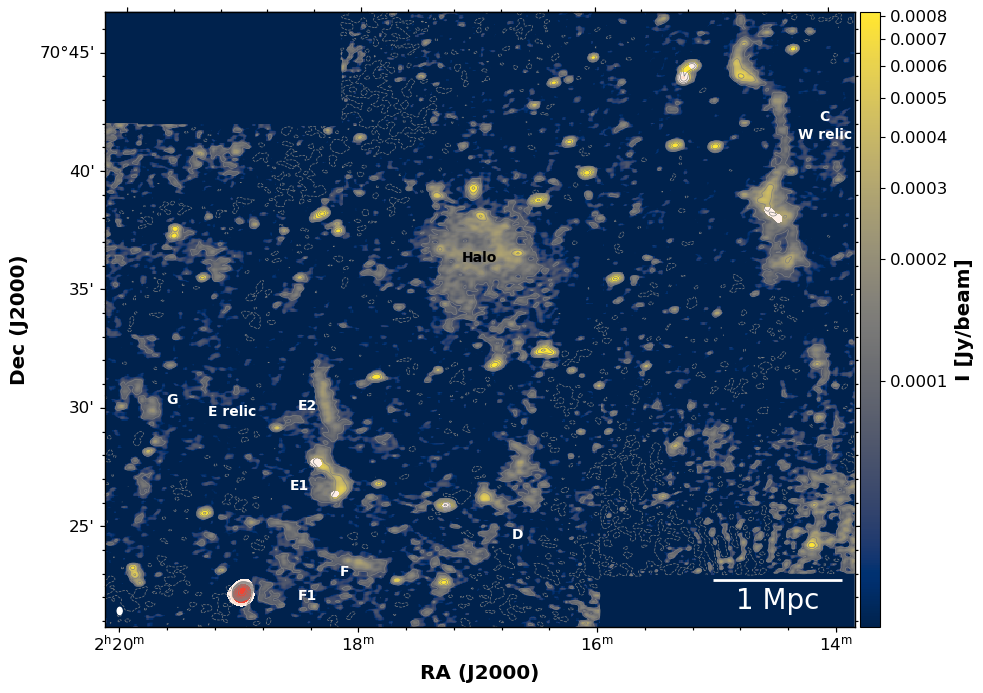}
    \caption{The galaxy cluster 8C0212. The blue-yellow color scale shows the full-resolution total intensity image at the central frequency of 1.5 GHz. Contours start at at $3\sigma$ (with $\sigma$ = 25 $\mu$Jy/beam) and increase by a factor of 2. The dashed contour shows the $-3\sigma$ level. Filled white-red contours show the polarized intensity image integrated over the full band with [6,12,24,96]$\sigma_{QU}$ levels. Beam size and $\sigma_{QU}$ values are listed in Tab.~\ref{tab:pol}. Diffuse sources are labeled following \citet{Hoang21}.}
    \label{fig:8C0212_nice}
\end{figure*}

We corrected the RM values for the Galactic foreground. We computed the median Galactic RM (GRM) in a 1 degree diameter circle around the galaxy cluster from \citet{Huts21}. GRM values are listed in Tab.~\ref{tab:clusters}. Finally we subtracted the GRM value from our RM maps to obtain the residual RM (RRM). In the following, we will consider RRM values of extra-galactic origin. Since the RM dispersion within 1 degree computed from the \citet{Huts21} map is lower than $6 \radm$ for the four considered clusters, the residual Galactic contribution on the angular size of radio relics (few arcminutes) should be well below this value.

In the following Section, we will quote only observed RRM values. This values can differ from the intrinsic RM value in the rest-frame of the Faraday screen due to the cosmological expansion. This difference can be particularly important for high-redshift sources \citep[see e.g.,][]{Carretti22}. Assuming that the the whole extra-galactic Faraday rotation occurs at the source redshift, i.e. in the ICM of the galaxy cluster, the rest-frame RRM is RRM$(1+$z$)^2$, with the $z$ of the cluster. In Sec.~\ref{sec:discuss} we will consider rest-frame RRM values.

\section{Results}
\label{sec:result}

In this Section we show total intensity, polarization, and RRM images obtained for the four galaxy clusters. For 8C0212 and A3365 we created total intensity composite images of the two separate pointings only for visualization. 

The polarized intensity images integrated over the full band and masked at 6$\sigma_{QU}$ are shown with filled contours on top of the total intensity images, in order to show the regions where polarized emission was detected. With respect to polarized intensity images obtained from RM synthesis, these images show much smoother detected regions since the masking is made with an average rms value while for the RM synthesis we masked on a per-pixel basis. Furthermore, they often have a lower rms noise level (see Tab.~\ref{tab:pol}) because the RM synthesis introduces additional noise that is collected in the Faraday spectrum. On the other hand, full-band integrated images suffer from in-band depolarization and are therefore less accurate for polarization measurements. Hence, we used them only for better visualization.

We detected extended polarized emission only for two out of eight radio relics, namely the eastern relic of Abell 3365 and the southern relic of PLCK287. For other three relics, i.e. the two relics in 8C0212 and the southern relic of ZwCl2341, we only detected few patches of polarized emission. The remaining three relics are unpolarized up to our detection threshold.

\subsection{8C 0212+703 (ClG 0217+70)}
\label{sec:8C0212results}

The galaxy cluster 8C0212 shows few patches of polarized emission at the position of the E1 source and of the brightest part of the western relic (see Fig.~\ref{fig:8C0212_nice}).

\begin{figure*}[h]     
    \centering
	\includegraphics[width=0.32\textwidth]{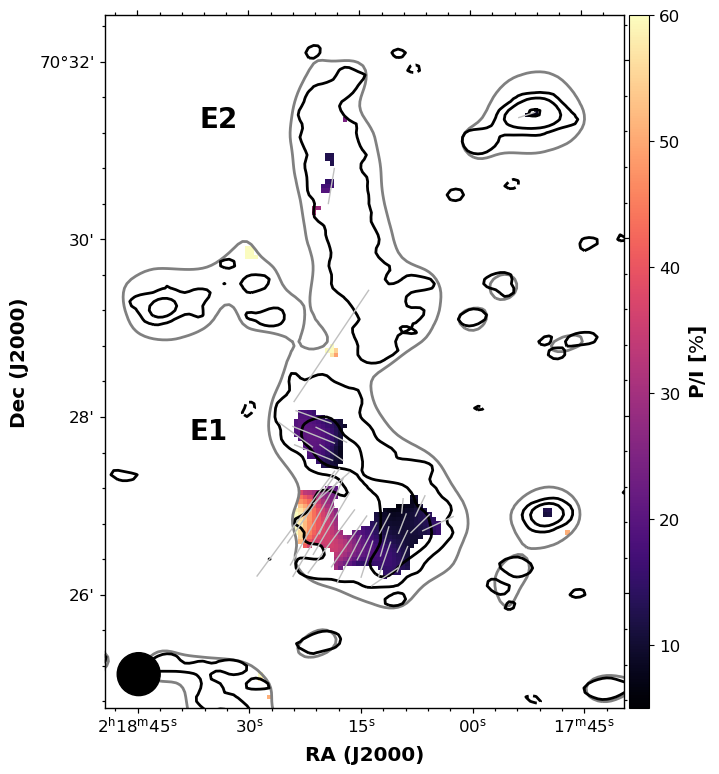}
	\includegraphics[width=0.33\textwidth]{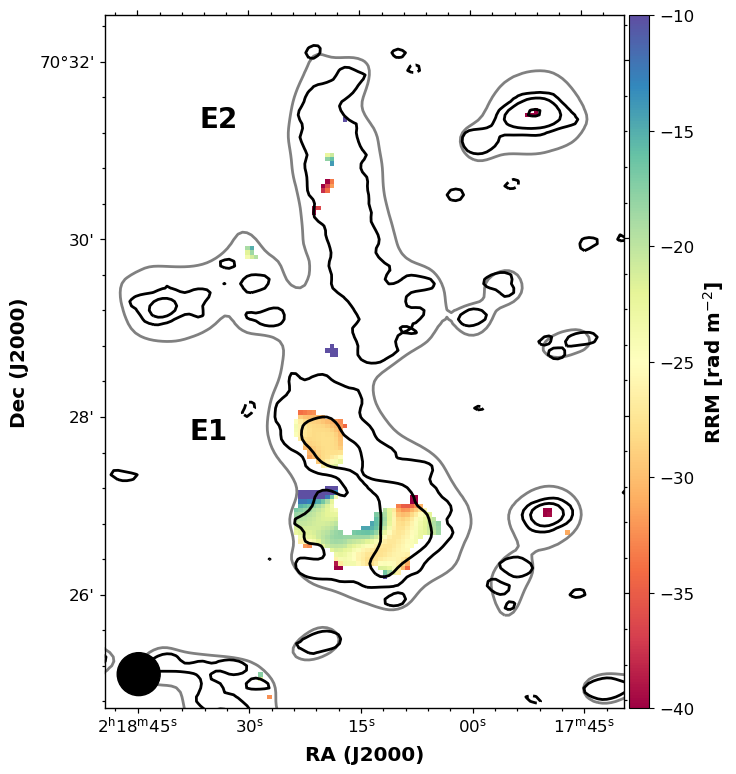}
    \caption{Fractional polarization and residual RM (i.e., corrected for Galactic Faraday rotation) images of source E in the eastern side of 8C0212. In the left-hand panel, gray vectors show the magnetic field direction and their length is proportional to the fractional polarization value. Black contours are [-3,3,12]$\sigma_I$ of the high-resolution total intensity image while the gray contour shows the 3$\sigma_I$ of the low-resolution total intensity image used to compute the fractional polarization. Only pixels above the $6\sigma_{\widetilde{Q}\widetilde{U}}$ are shown. Rms noise levels and beam sizes are listed in Tab.~\ref{tab:pol}.}
    \label{fig:8C0212_polE}
\end{figure*}

\begin{figure*}[h]
    \centering
	\includegraphics[width=0.37\textwidth]{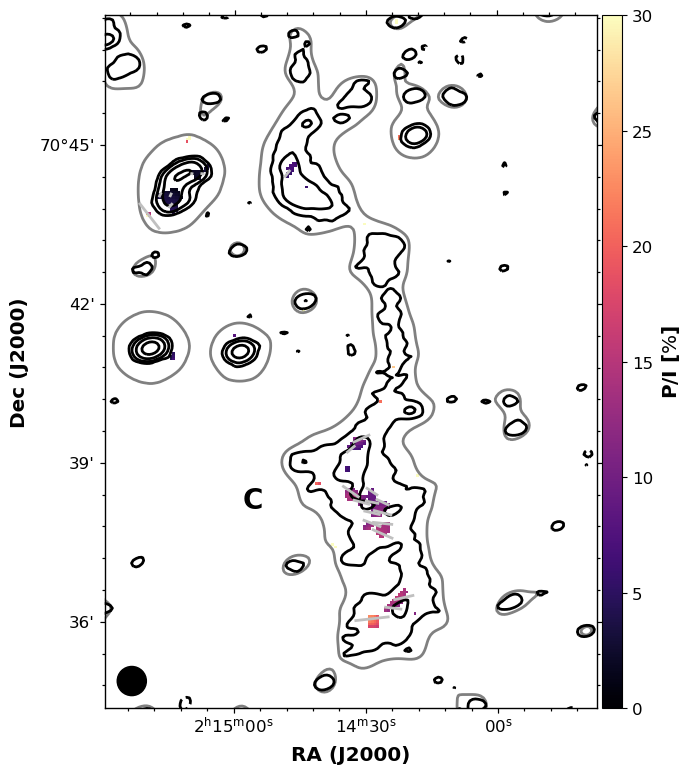}
	\includegraphics[width=0.38\textwidth]{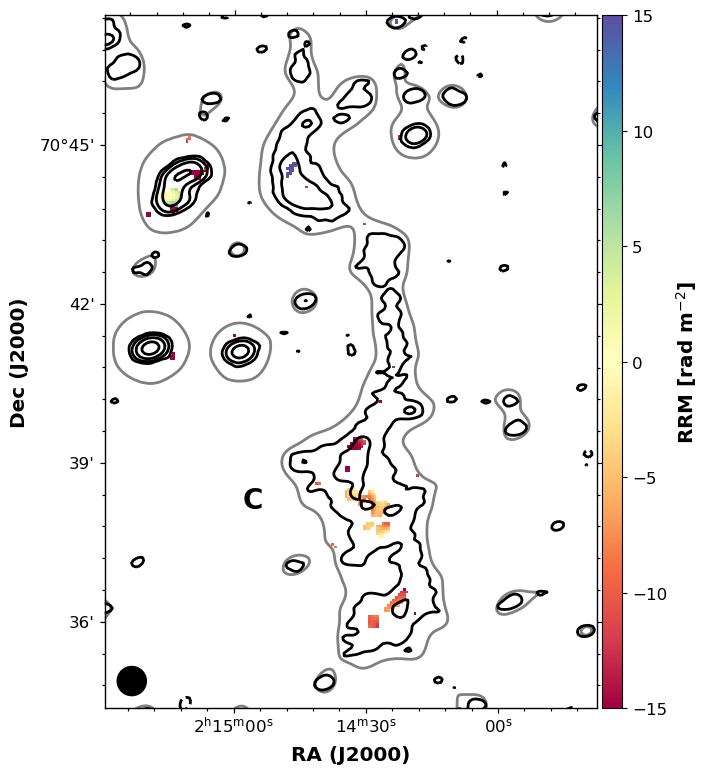}
    \caption{Fractional polarization and residual RM (i.e., corrected for Galactic Faraday rotation) images the western relic of 8C0212 (source C). In the left-hand panel, gray vectors show the magnetic field direction and their length is proportional to the fractional polarization value. Black contours are [-3,3,12,48,192]$\sigma_I$ of the high-resolution total intensity image while the gray contour shows the 3$\sigma_I$ of the low-resolution total intensity image used to compute the fractional polarization. Only pixels above the $6\sigma_{\widetilde{Q}\widetilde{U}}$ are shown. Rms noise levels and beam sizes are listed in Tab.~\ref{tab:pol}.}
    \label{fig:8C0212_polW}
\end{figure*}

Source E is close, at least in projection, to the eastern radio relic and was distinguished in two components (E1 and E2) by \citet{Hoang21} on the basis of their morphology. E1 has a double lobed structure, typical of a Fanaroff-Riley type I radio galaxy, with an optical counterpart in Pan-STARRS. The southern lobe is bent towards the E direction while the northern one is bent towards NW where it merge with E2 (see also Fig.~\ref{fig:8C0212_polE}). E2 has an elongated structure and spectral index steepening towards the cluster center \citep{Hoang21}. Spectral index variations between E1 and E2 suggest a possible shock-induced re-acceleration of the fossil plasma ejected by the southern AGN.

\begin{figure*}[h]
    \centering
	\includegraphics[width=0.9\textwidth]{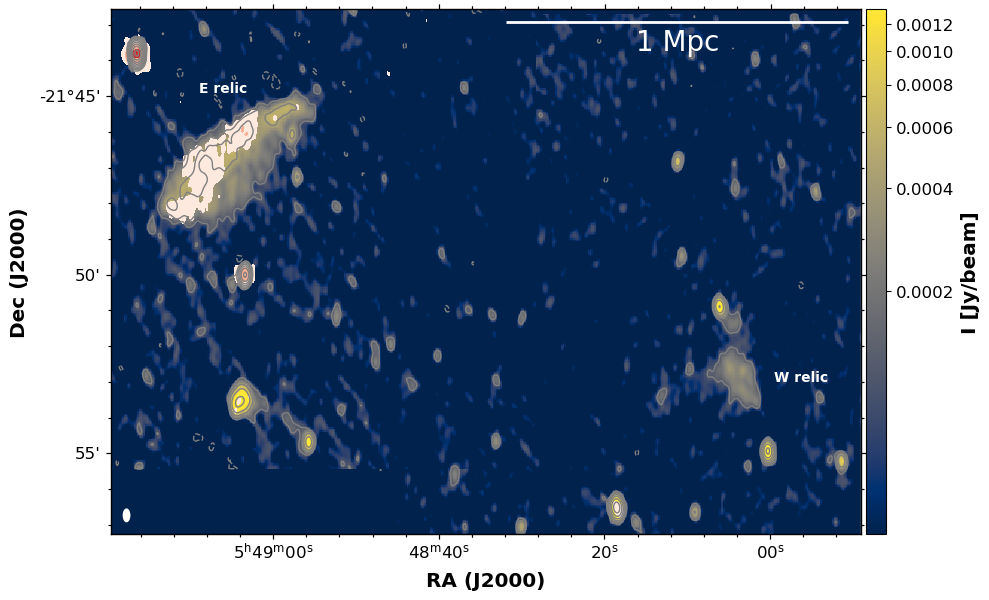}
    \caption{The galaxy cluster A3365. The blue-yellow color scale shows the full-resolution total intensity image at the central frequency of 1.5 GHz. Contours start at at $3\sigma$ (with $\sigma$ = 40 $\mu$Jy/beam) and increase by a factor of 2. The dashed contour shows the $-3\sigma$ level. Filled white-red contours show the polarized intensity image integrated over the full band with [6,12,24,48]$\sigma_{QU}$ levels. Beam size and $\sigma_{QU}$ values are listed in Tab.~\ref{tab:pol}. The two relics are labeled.}
    \label{fig:A3365_nice}
\end{figure*}

\begin{figure*}[h]
    \centering
	\includegraphics[width=0.42\textwidth]{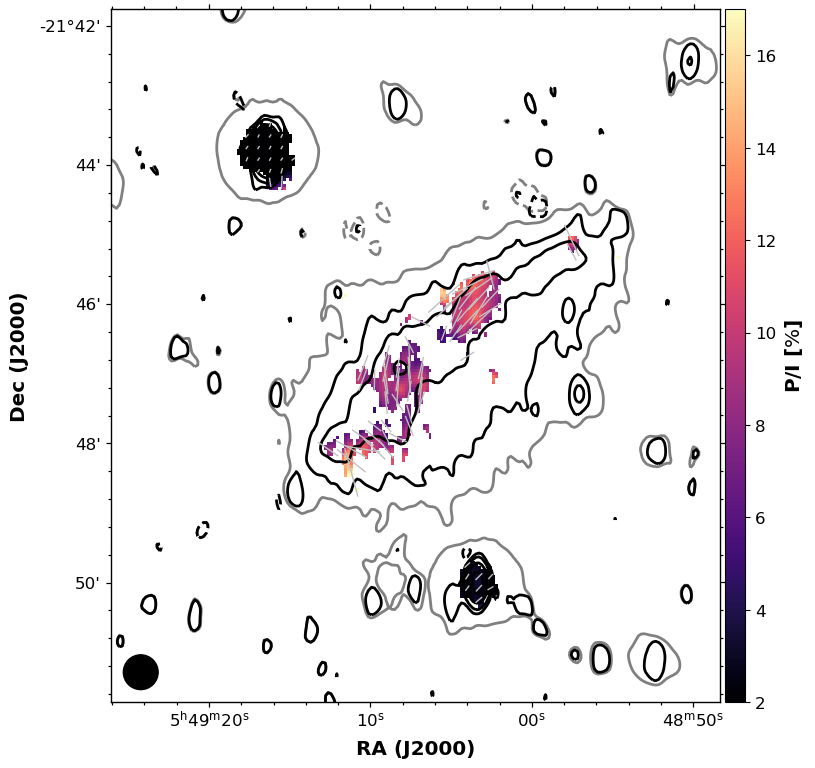}
	\includegraphics[width=0.43\textwidth]{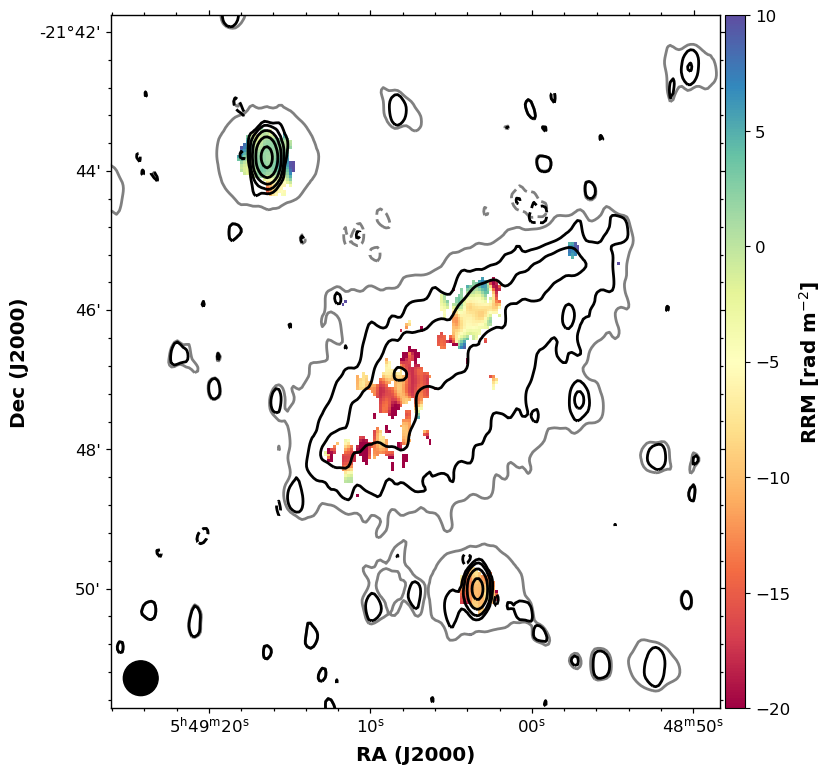}
    \caption{Fractional polarization and residual RM (i.e., corrected for the Galactic Faraday rotation) images of the eastern relic of A3365. In the left-hand panel, gray vectors show the magnetic field direction and their length is proportional to the fractional polarization value. Black contours are [-3,3,12,48,192,768]$\sigma_I$ of the high-resolution total intensity image while the gray contour shows the $\pm3\sigma_I$ of the low-resolution total intensity image used to compute the fractional polarization. Only pixels above the $6\sigma_{\widetilde{Q}\widetilde{U}}$ are shown. Rms noise levels and beam sizes are listed in Tab.~\ref{tab:pol}.}
    \label{fig:A3365_pol}
\end{figure*}

We detected polarized emission from the lobes of the radio galaxy (source E1, see Fig.~\ref{fig:8C0212_polE}). The average fractional polarization is $16\pm2 \ \%$ and $17\pm2 \ \%$ in the northern and southern lobes, respectively, but it reaches values of $60 \ \%$ in the eastern extension of the southern lobe. Magnetic field vectors are aligned with the main axis of the northern lobe while they get aligned in the perpendicular direction towards the eastern extension up to the edge, where the fractional polarization reaches its maximum. Interestingly, at this position the spectral index becomes flatter, as reported by \citet{Hoang21}. This suggests that in this region a physical process is simultaneously accelerating particles and aligning the plane-of-the-sky magnetic field components. The presence of a shock wave, as already proposed by \citet{Hoang21} to explain the spectral index properties of E2, would furnish a good explanation also for the eastern extension of the southern lobe. However, without deeper X-ray images able to detect the presence of a shock, this remains an hypothesis. Furthermore, the fact that E2 is not detected in polarization with an upper limit of $13 \ \%$ challenges this interpretation. 

The median RRM in the northern lobe is $-28\radm$ with a standard deviation, $\sigma_\textrm{RM}$, of $2\radm$. The brightest part of the southern lobe has similar values of RRM, while the RRM increases to a median value of $-20\radm$ in the eastern extension (see Fig.~\ref{fig:8C0212_polE}, right panel). This difference could be attributed to both projection effects (with the brightest part of the lobe being closer), magnetic field strength and/or thermal electron density variations or to magnetic field reversals.

We did not detect polarization from sources D, F, and G. These sources are very faint and only partially detected by our observations. The upper limit to their fractional polarization is: $28 \ \%$, $22 \ \%$ and $26 \ \%$, respectively.

We detected few patches of polarized emission arising from the western radio relic (source C, see Fig.~\ref{fig:8C0212_polW}). The average fractional polarization here is $12\pm2 \ \%$ with a maximum values of $\sim23 \ \%$  while the RRM has a median value of $-7\radm$ and $\sigma_\textrm{RM} = 10\radm$. Magnetic field vectors are broadly aligned with the main axis of the radio emission. Overall, the detection of polarized emission confirm the identification of the C source with a radio relic. However, considering the elongated shape of this radio relic and its peripheral position ($\sim2.4$ Mpc from the cluster center), we would expect to detect higher fractional polarization values. This will be discussed in Sec.~\ref{sec:depol}.

\subsection{Abell 3365}
\label{sec:A3365results}

The elongated eastern radio relic of A3365 shows extended polarized emission while the western relic remains undetected in polarization (see Fig.~\ref{fig:A3365_nice}).

The zoomed view of the eastern relic is shown in Fig.~\ref{fig:A3365_pol}. The polarized emission is detected from the brightest region of the relic, but is more patchy with respect to the total intensity emission. The average fractional polarization in the detected regions is $9.0\pm0.8 \ \%$, reaching a maximum value of $18 \ \%$. Polarization vectors are parallel to the main axis of the relic only in the northern part, while they bent and become perpendicular towards the south. The median RRM is $-11\radm$ and $\sigma_\textrm{RM} = 11\radm$. The RM values are more scattered in the northern part of the relic. 

The western relic is much fainter than the eastern one and it has not the classical arc-like shape. The upper limit to its fractional polarization is $8 \ \%$. Its low fractional polarization could be due to projection effects. This will be further discussed in Sec.~\ref{sec:depol}.

\subsection{PLCK G287.0+32.9 (PSZ2 G286.98+32.90)}
\label{sec:PLCK287results}

\begin{figure*}[h]
    \centering
	\includegraphics[width=0.8\textwidth]{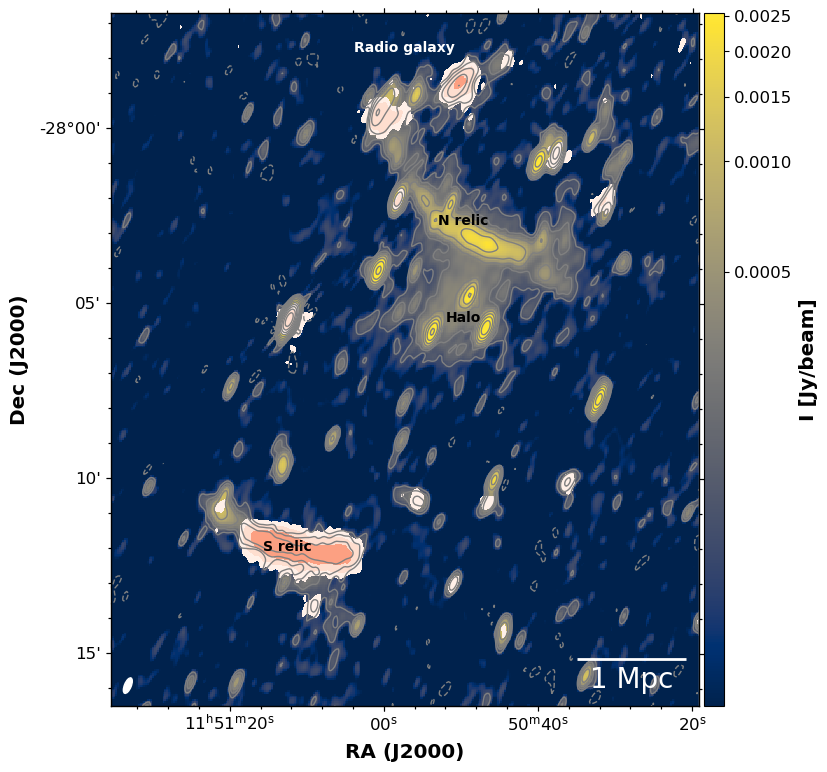}
    \caption{The galaxy cluster PLCK287. The blue-yellow color scale shows the full-resolution total intensity image at the central frequency of 1.5 GHz. Contours start at at $3\sigma$ (with $\sigma$ = 33 $\mu$Jy/beam) and increase by a factor of 2. The dashed contour shows the $-3\sigma$ level. Filled white-red contours show the polarized intensity image integrated over the full band with [6,12,24,48,96]$\sigma_{QU}$ levels. Beam size and $\sigma_{QU}$ values are listed in Tab.~\ref{tab:pol}. Notable diffuse radio sources are labeled.}
    \label{fig:PLCK287_nice}
\end{figure*}

\begin{figure*}[h]
    \centering
	\includegraphics[width=0.47\textwidth]{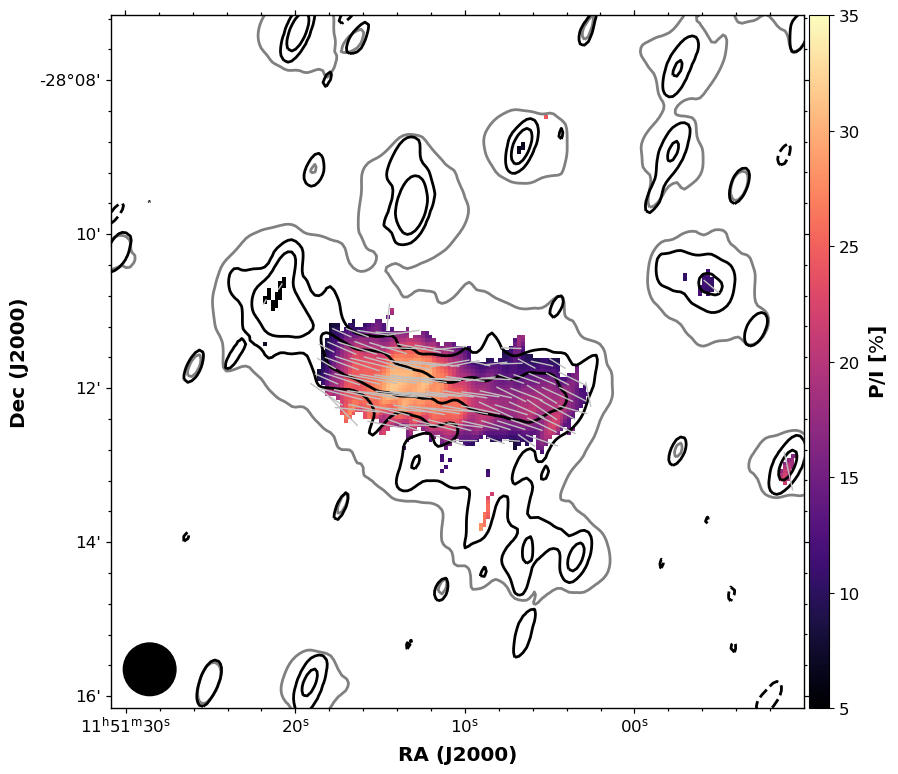}
	\includegraphics[width=0.48\textwidth]{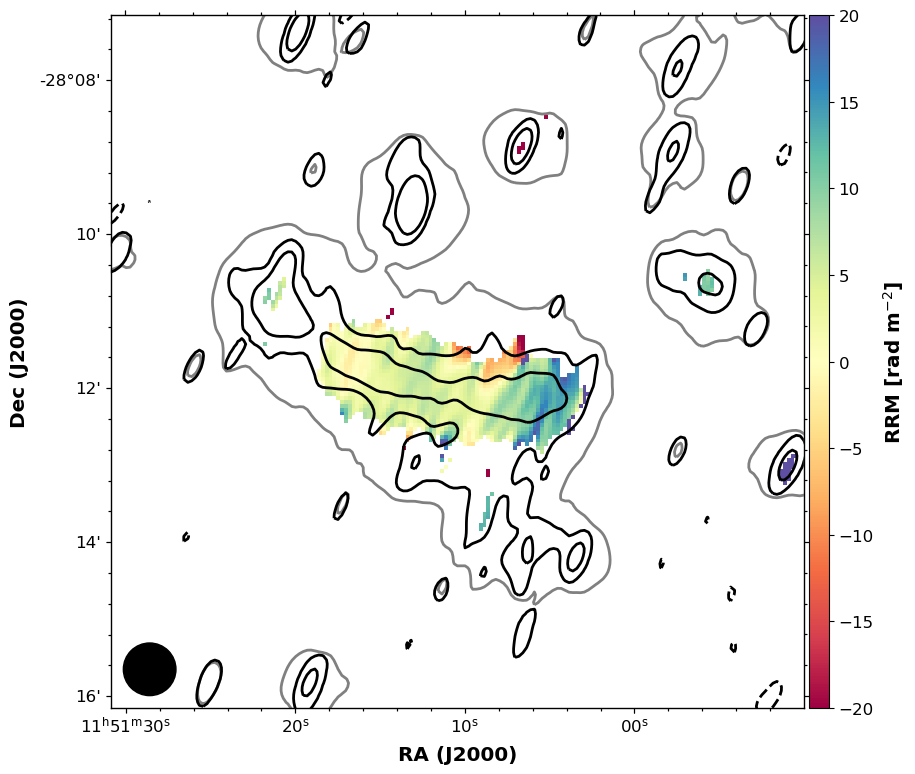}
    \caption{Fractional polarization and residual RM (i.e., corrected for the Galactic Faraday rotation) images of the southern relic of PLCK287. In the left-hand panel, gray vectors show the magnetic field direction and their length is proportional to the fractional polarization value. Black contours are [-3,3,12,48,192]$\sigma_I$ of the high-resolution total intensity image while the gray contour shows the $3\sigma_I$ of the low-resolution total intensity image used to compute the fractional polarization. Only pixels above the $6\sigma_{\widetilde{Q}\widetilde{U}}$ are shown. Rms noise levels and beam sizes are listed in Tab.~\ref{tab:pol}.}
    \label{fig:PLCK287_pol}
\end{figure*}

We detected diffuse polarized emission from the southern radio relic in PLCK287 (see Fig.~\ref{fig:PLCK287_nice} and Fig.~\ref{fig:PLCK287_pol}). Polarization vectors are well aligned with the main axis of this relic and the fractional polarization reaches the $31 \ \%$ with an average value of $20\pm1 \ \%$. We have also detected few patches of polarized emission arising from the southern extension of this relic, previously noticed by \citet{Bonafede14}, thus supporting its connection to the radio relic. The median RRM at the southern relic is $5\radm$ with $\sigma_\textrm{RM} = 8\radm$. We observed a gradient of decreasing RRM going from the western side of the relic to the east, where the RRM approaches zero (see Fig.~\ref{fig:PLCK287_pol}). This behavior suggests a possible inclination of the relic on the plane of the sky, with the western part lying deeper in the ICM and experiencing more Faraday rotation. Galactic RM variation across the sources on arcminutes-scales cannot be excluded.

We also detected the polarized lobes of the radio galaxy in the north of the cluster (see Fig.~\ref{fig:PLCK287_nice}). The southern lobe of the radio galaxy is connected to the northern radio relic but polarization was not detected neither from the relic nor from the radio bridge between the two sources. The upper limit to the fractional polarization of the northern radio relic is $0.8 \ \%$. Indeed this relic is very bright and if any polarization was present we should have detected it. We know from \citet{Bonafede14} that the northern relic is close in projection to the galaxy cluster center and to its X-ray emission peak. It is possible that it is located in behind the bulk of the ICM and that the external Faraday dispersion totally depolarize the signal within our resolution beam ($\sim220$ kpc, see also Sec.~\ref{sec:depol}). The lobes of the radio galaxy are far away from the cluster center and we were able to detect their polarized emission. \citet{Bonafede14} did not find the optical counterpart of this source using the Wide Field Imager (WFI). We also searched for a counterpart with redshift estimate in the NED database, without success. However, the RRM of the radio galaxy lobes are similar to those of the nearest galaxy which is a confirmed galaxy cluster member (being $19\radm$, $25\radm$ and $32\radm$ the median RRM of the north-western lobe, south-eastern lobe and of the nearest galaxy cluster member detected in polarization, respectively, and $21\radm$, $19\radm$ and $16\radm$, their RM dispersion). This support the idea that the radio galaxy is in the same environment of the radio relic.

\subsection{ZwCl 2341.1+0000}
\label{ZwCl2341results}

\begin{figure*}[h]
    \centering
	\includegraphics[width=0.7\textwidth]{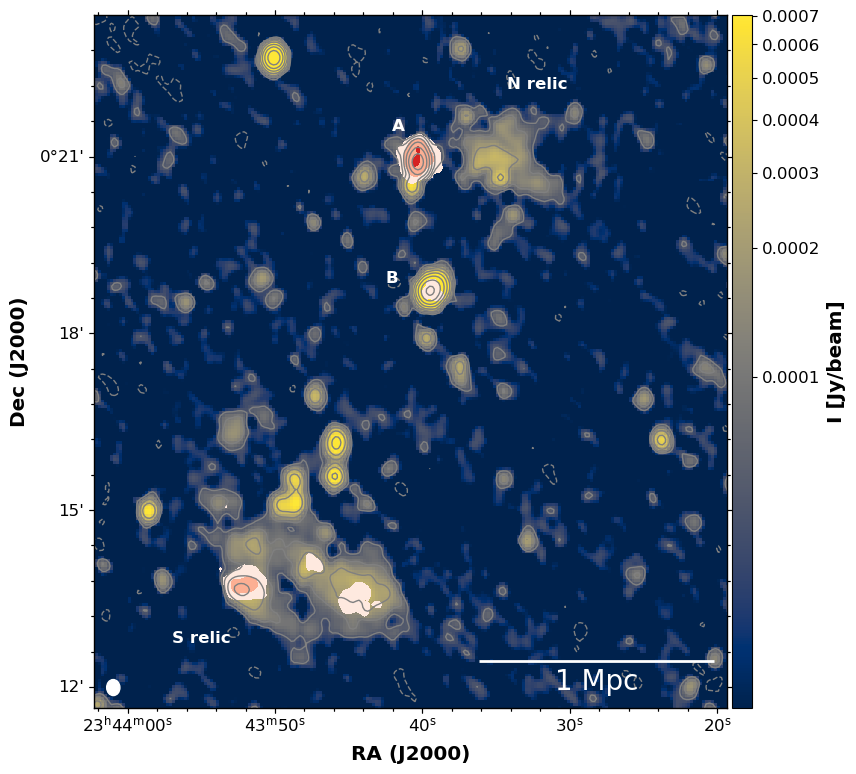}
    \caption{The galaxy cluster ZwCl2341. The blue-yellow color scale shows the full-resolution total intensity image at the central frequency of 1.5 GHz. Contours start at at $3\sigma$ (with $\sigma$ = 17 $\mu$Jy/beam) and increase by a factor of 2. The dashed contour shows the $-3\sigma$ level. Filled white-red contours show the polarized intensity image integrated over the full band with [6,12,24,48]$\sigma_{QU}$ levels. Beam size and $\sigma_{QU}$ values are listed in Tab.~\ref{tab:pol}. The two radio relics and two radio galaxies detected in polarization are labeled.}
    \label{fig:ZwCl2341_nice}
\end{figure*}

Our total intensity image of ZwCl2341 (Fig.~\ref{fig:ZwCl2341_nice}) is similar to those recently presented by \citet{Parekh22} and obtained with the MeerKAT radio telescope at 1.28 GHz. However, we do not confirm their marginal detection of extended emission in the galaxy cluster center.

\begin{figure*}[h]
    \centering
	\includegraphics[width=0.37\textwidth]{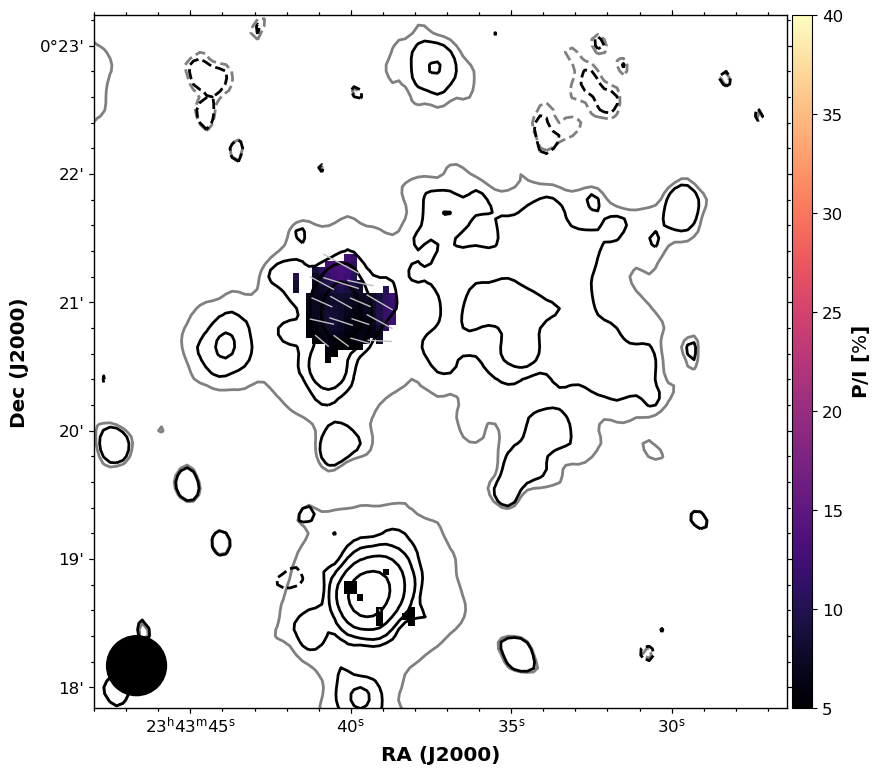}
	\includegraphics[width=0.39\textwidth]{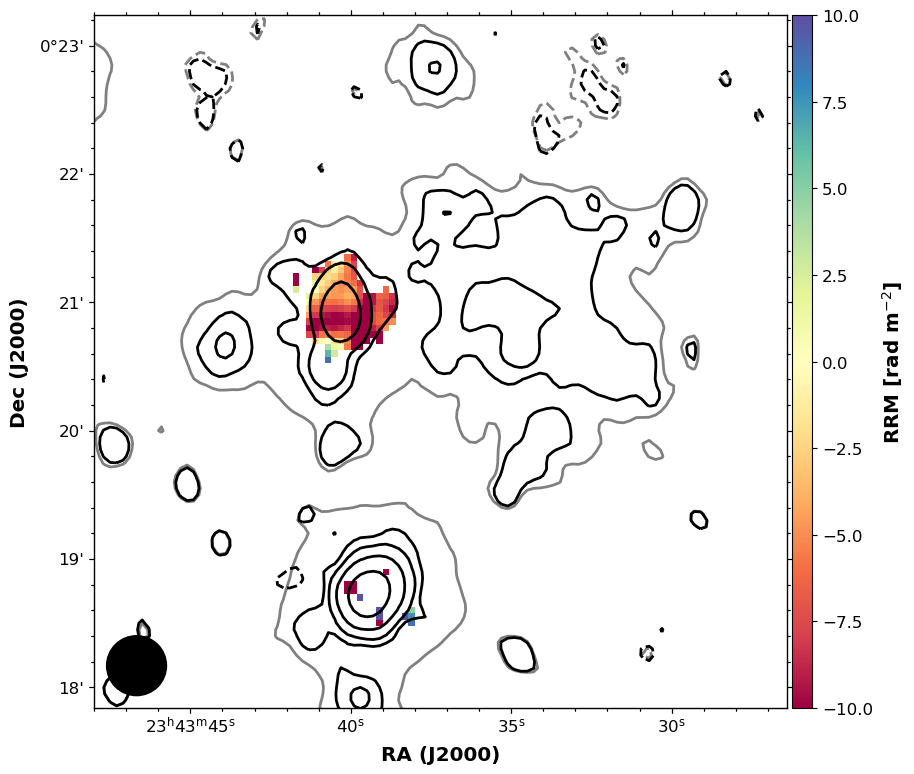}
    \caption{Fractional polarization and residual RM (i.e., corrected for the Galactic Faraday rotation) images centered on the northern relic of ZwCl2341. In the left-hand panel, gray vectors show the magnetic field direction and their length is proportional to the fractional polarization value. Black contours are [-3,3,12,48]$\sigma_I$ of the high-resolution total intensity image while the gray contour shows the $\pm3\sigma_I$ of the low-resolution total intensity image used to compute the fractional polarization. Only pixels above the $6\sigma_{\widetilde{Q}\widetilde{U}}$ are shown. Rms noise levels and beam sizes are listed in Tab.~\ref{tab:pol}. The only detected sources in polarization are two cluster's radio galaxies.}
    \label{fig:ZwCl2341_pol_N}
\end{figure*}

The northern radio relic of the ZwCl2341 system is unpolarized in our observations. Its polarization fraction has an upper limit of $5 \ \%$. A zoomed view of this relic and of the surrounding sources is shown in Fig.~\ref{fig:ZwCl2341_pol_N}. A cluster radio galaxy to the east of the relic (source A in Fig.~\ref{fig:ZwCl2341_nice}), possibly classified as an head-tail source by \citet{vanWeeren09}, has an average fractional polarization of $7.7\pm0.7 \ \%$ with a maximum value of $13 \ \%$. Another polarized source (source B) is observed towards the cluster center and its high RM dispersion ($94 \radm$) suggests that it is located deeper in the ICM. It is in fact a member of this galaxy cluster \citep{vanWeeren09}.

\begin{figure*}[h]
    \centering
	\includegraphics[width=0.44\textwidth]{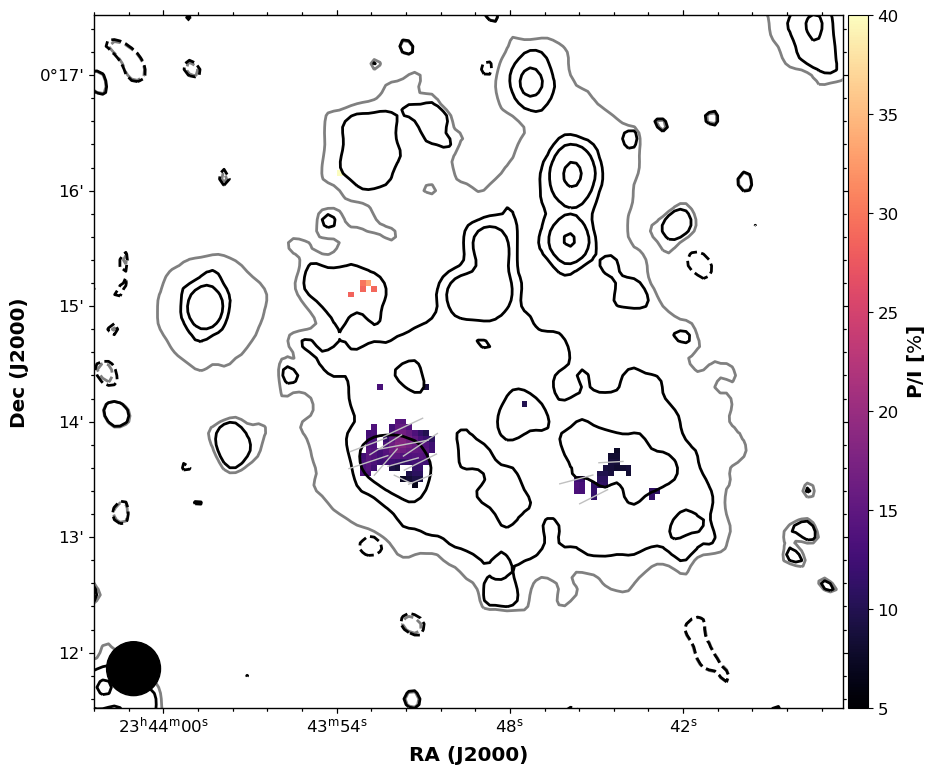}
	\includegraphics[width=0.46\textwidth]{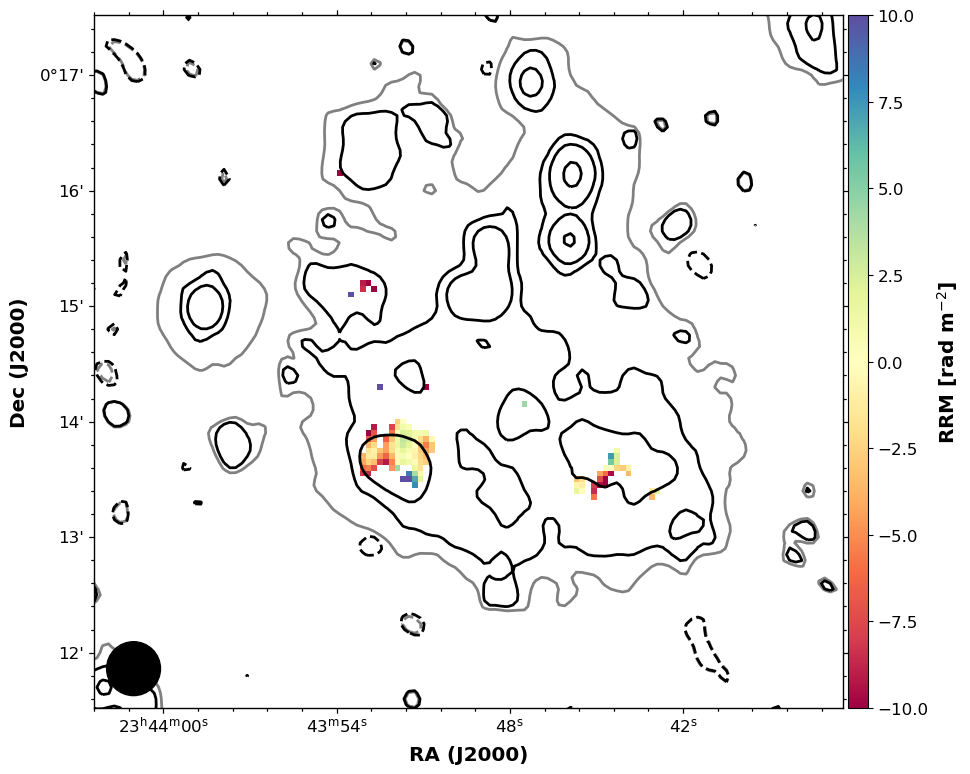}
    \caption{Fractional polarization and residual RM (i.e., corrected for the Galactic Faraday rotation) images of the southern relic of ZwCl2341. In the left-hand panel, gray vectors show the magnetic field direction and their length is proportional to the fractional polarization value. Black contours are [-3,3,12,48]$\sigma_I$ of the high-resolution total intensity image while the gray contour shows the $\pm3\sigma_I$ of the low-resolution total intensity image used to compute the fractional polarization. Only pixels above the $6\sigma_{\widetilde{Q}\widetilde{U}}$ are shown. Rms noise levels and beam sizes are listed in Tab.~\ref{tab:pol}.}
    \label{fig:ZwCl2341_pol_S}
\end{figure*}

The southern relic shows few patches of polarized emission with average fractional polarization of $13\pm2 \ \%$ (see Fig.~\ref{fig:ZwCl2341_pol_S}). There are few pixels with fractional polarization reaching the $33 \ \%$ but the polarized emission is concentrated in the brightest relic region in the south, which has a roundish morphology and a maximum fractional polarization of $18 \ \%$. We checked that this is not a point-source since it disappears increasing the image resolution. The median RRM is $-2\radm$ and $\sigma_\text{RM} = 22\radm$. 

\citet{Giovannini10} reported the detection of polarized emission all over the two relics and also in the region between them using VLA low resolution images ($83\arcsec\times75\arcsec$, $\sim 330$ kpc at the cluster's redshift) at 1.4 GHz. They found $15 \ \%$ fractional polarization in the northern relic and $8 \ \%$ in the southern one. While the value found in the southern relic is consistent with our measurements (considering beam-depolarization), our non-detection of the northern relic is at odds with their findings. We suggest that this is due to the contamination from the polarized emission of the head-tail radio galaxy to the east of the relic (source A in Fig.~\ref{fig:ZwCl2341_nice}) which could be under-subtracted in low-resolution images. In our images the fractional polarization of the AGN reaches the $13 \ \%$ and the magnetic field direction at the source is consistent with the one obtained by \citet{Giovannini10}. Also, we checked that using only the D configuration and integrating over the full band we can recover polarized emission at $5 \ \%$ level from the northern relic, but this emission is not present after the RM synthesis due to the low signal-to-noise. \citet{Benson17} measured the fractional polarization of the two relics in the $2-4$ GHz band obtaining $5 \ \%$ for the northern relic and $8 \ \%$ for the southern one. In this case, the value found in the northern relic is consistent with our upper limit while the southern relic show a lower polarization fraction. This could be due to the fact that we only detected polarized emission from the brightest and most polarized regions of the relic. The image produced in the $2-4$ GHz band by \citet{Benson17} shows polarized emission all over the relic possibly catching also low fractional polarization regions. A polarization study performed combining L- and S-band measurements would be necessary to deeply investigate this difference and to unveil possible complex Faraday structures in this relic.

However, given the disturbed structure of this cluster, the non-detection of surface brightness edges from deep \textit{Chandra} observations \citep{Zhang21} and the possible presence of a secondary merger along the line-of-sight \citep{Golovich19b}, it is very likely that these relics are not seen edge-on and that projection effects play a role in determining their depolarization fraction. This is also supported by the RM dispersion of the southern relic which is the highest observed in our sample.

\section{Discussion}
\label{sec:discuss}

In this Section we compare our results with literature information about double radio relics and with state-of-the-art of MHD simulations.

In Tab.~\ref{tab:double_all}, we made a compilation of all known double radio relics in the literature. For each of them we listed, when available, fractional polarization (average and maximum) and Faraday rotation (average/median subtracted from the Galactic foreground and dispersion). We computed here RRM and $\sigma_\textrm{RM}$ values as measured in the source rest-frame, assuming that all the residual RM and RM dispersion are generated at the galaxy cluster's redshift and therefore we multiplied the observed values by $(1+z)^2$ (see also Sec.~\ref{sec:im}). This operation is needed to compare observational results with simulations although it is often neglected in the literature. We considered only observations at 1.4 GHz in order to compare with our observations (with the only exception of the El Gordo galaxy cluster for which 1.4 GHz observations are not available).

Our collection resulted in 22 double radio relics systems. We found polarization information for 15 clusters but Faraday rotation measurements are available only for 9 of them. Therefore, with this work we have almost doubled the number of double relics galaxy clusters with RM information. The results reported in the table highlight that a lot of information is still missing in order to have a complete view about the polarization properties of double radio relic systems.

\begin{table*}[!h]
    \centering
	\caption{Compilation of double radio relic galaxy clusters. Column 1: name of the cluster. A subscript letter specifies the relic; Column 2: redshift, $z$, retrieved from NED with the exception of 8C 0212+703 for which an updated redshift is provided by \citet{Zhang20}, and Abell 2146 for which the most updated redshift is in \citet{White15}; Column 3: average fractional polarization at 1.4 GHz; Column 4: maximum fractional polarization at 1.4 GHz; Column 5: median or average residual RM, already corrected for the Galactic contribution and for cosmological effects; Column 6: rest-frame corrected RM dispersion. Except for the northern relic in CIZA J2242.8+5301, for which QU-fitting was used, this value is computed over the relic area. For some relics we listed the total range of measured RM values included in squared brackets since the $\sigma_\textrm{RM}$ is not reported; Column 7: observing beam of polarization observations; Column 8: physical resolution of the polarization observations; Column 9: projected distance of the relic from the X-ray centroid of the hosting galaxy cluster; Column 10: references for polarization observations, as for the legend. From this double radio relics list we excluded Abell 548b \citep{Feretti06}, CIZA J0107.7+5408 \citep{Randall16}, MACS J0025.4-1222 \citep{Riseley17}, SPT-CL J0245-5302 \citep{Zheng18}, RXC J2351.0-1934 \citep{Duchesne21b}, and MACS J0417.5-1154, MCXC J0232.2-4420, MCXC J0516.6-5430 \citep{Knowles22} for which the characterization as double radio relics is still uncertain.}
	\label{tab:double_all}
	\resizebox{\linewidth}{!}{
	\begin{tabular}{lccccccccc}
		\hline
		\hline
		Cluster$_\textrm{relic}$ & $z$ & <p>$_{\rm 1.4 GHz}$ & max. p$_{\rm 1.4 GHz}$  & RRM$(1+z)^2$ & $\sigma_\textrm{RM}(1+z)^2$ & Obs. Beam & Physical resolution & d & References \\
		         &    & [$\%$] & [$\%$] & [$\radm$] & [$\radm$] &   &  [kpc] &  [Mpc] &    \\
        \hline
        8C 0212+703$_E$ (ClG 0217+70)$^a$ & 0.180 & <22 & -- & -- & -- & $29\arcsec\times29\arcsec$  & 89 & 2.5 &  tw\\
        8C 0212+703$_W$ (ClG 0217+70) & 0.180 & 12$\pm$2 & 23 & -10 & 14 & $33\arcsec\times33\arcsec$ & 101 & 2.4 & tw\\
        \hline
        Abell 1240$_N$  &  0.195 & 26 & 70 & -- & -- & $18\arcsec\times18\arcsec$ & 59 & 0.7 & Bo09 \\
        Abell 1240$_S$    &  0.195 & 29 & 70 & -- & -- & $18\arcsec\times18\arcsec$ & 59 & 1.1 & Bo09 \\
        \hline
        Abell 2146$_N$   &  0.232 & -- & -- & -- & -- & -- & -- & 0.45 &  \\
        Abell 2146$_S$   &  0.232 & -- & -- & -- & -- & -- & -- & 0.2 & \\
        \hline
        Abell 2345$_E$  & 0.176 & 18$\pm$1 & 70 & -0.3 & 8 & $30.5\arcsec\times30.5\arcsec$ & 92 & 0.9 & S21 \\
        Abell 2345$_W$  & 0.176 & 12.6$\pm$0.9 & 70 & -7 & 18 & $30.5\arcsec\times30.5\arcsec$ & 92 & 1.0 & S21  \\
        \hline
        Abell 3186$_{NW}$ (MCXC J0352.4-7401) &  0.127 & -- & -- & -- & -- & -- & -- & 1.5 &  \\
        Abell 3186$_{SE}$ (MCXC J0352.4-7401) &  0.127 & -- & -- & -- & -- & -- & -- & 1.2 & \\ 
        \hline
        Abell 3365$_E$  &  0.093 & 9.0$\pm$0.8 & 18 & -13 & 13 & $30\arcsec\times30\arcsec$ & 52 & 1.0 & tw  \\
        Abell 3365$_W$ &  0.093 & <8 & -- & -- & -- & $31\arcsec\times31\arcsec$  & 54 & 1.0 & tw \\
        \hline
        Abell 3376$_E$    &  0.046 & -- & 30 & -- & -- & $37\arcsec\times25\arcsec$ & 34 & 0.4 & Ka12  \\
        Abell 3376$_W$   &  0.046 & -- & 20 & -- & -- & $38\arcsec\times26\arcsec$ & 34 & 1.6 & Ka12 \\ 
        \hline
        Abell 3667$_{NW}$    &  0.055 & 20 & 70 & -- & [0-11] & $10\arcsec\times10\arcsec$ & 11 & 1.5 & dG22  \\
        Abell 3667$_{SE}$ &  0.055 & 25 & 50 & -- & [0-44] & $10\arcsec\times10\arcsec$ & 11 & 1.1 & dG22 \\
        \hline
        Abell 521$_{NW}$    &  0.247 & -- & -- & -- & -- & -- & -- & 1.0 &   \\
        Abell 521$_{SE}$    &  0.247 & -- & -- & -- & -- & -- & -- & 0.75 &  \\ 
        \hline
        ACT-CL J0102-4915$_{NW}$ (El Gordo)$^b$ &  0.87 & 33$\pm$1 & 67 & 38 & 21 & $11\arcsec\times11\arcsec$ & 86 & 1.4 & Li14 \\
        ACT-CL J0102-4915$_{E}$ (El Gordo)$^b$ &  0.87 & 33$\pm$3 & -- & -- & -- & $11\arcsec\times11\arcsec$ & 86 & 0.4 & Li14  \\
        \hline
        CIZA J2242.8+5301$_N$ (Sausage)  & 0.189 & 18 & 40 & -86$^c$ & 31 & $7\arcsec\times7\arcsec$ & 22 & 1.5 & DG21  \\
        CIZA J2242.8+5301$_S$ (Sausage) & 0.189 & 20 & 22 & -- & [-14-0] & $13\arcsec\times13\arcsec$ & 41 & 1.1 & DG21 \\ 
        \hline
        MACS J1752.0+4440$_{NE}$   &  0.366 & 20 & 40 & -- & -- & $17\arcsec\times12\arcsec$ & 87 & 1.3 & Bo12 \\
        MACS J1752.0+4440$_{SW}$    &  0.366 & 10 & 40 & -- & -- & $17\arcsec\times12\arcsec$ & 87 & 0.8 & Bo12 \\ 
        \hline
        PLCK G200.9-28.2$_{NE}$ & 0.22 & -- & -- & -- & -- & -- & -- & 0.6 & \\
        PLCK G200.9-28.2$_{SW}$ & 0.22 & -- & -- & -- & -- & -- & -- & 0.9 & \\
        \hline
        PLCK G287.0+32.9$_N$ (PSZ2 G286.98+32.90) & 0.39 & <0.8 & -- & -- & -- &$41\arcsec\times41\arcsec$ & 219 & 0.4 & tw \\
        PLCK G287.0+32.9$_S$ (PSZ2 G286.98+32.90) & 0.39 & 20$\pm$1 & 31 & 10 & 15 & $41\arcsec\times41\arcsec$ & 219 & 2.8 & tw \\
        \hline
        PSZ1 G096.89+24.17$_N$ (ZwCl 1856.8+6616) & 0.304 & -- & 60 & -- & -- & $10\arcsec\times9\arcsec$ & 45 & 0.75 & J21 \\
        PSZ1 G096.89+24.17$_S$ (ZwCl 1856.8+6616) & 0.304 & -- & 20 & -- & -- & $10\arcsec\times9\arcsec$ & 45 & 1.1 & J21 \\
        \hline
        PSZ1 G108.18-11.53$_{NE}$ & 0.335 & -- & 30 & -- & -- & $17\arcsec\times13\arcsec $ & 82 & 1.7 & dG15 \\
        PSZ1 G108.18-11.53$_{SW}$  & 0.335 & -- & 30 & -- & -- & $17\arcsec\times13\arcsec$ & 82 & 1.3 & dG15 \\
        \hline
        PSZ2 G233.68+36.14$_N$ & 0.345 & -- & -- & -- & -- & -- & -- & 0.7 & \\
        PSZ2 G233.68+36.14$_{SE}$ & 0.345 & -- & -- & -- & -- & -- & -- & 1.0 & \\
        \hline
        RXC J1314.4-2515$_E$ & 0.247 & 11 & 14 & -5 & 17 & $25\arcsec\times25\arcsec$ & 98 & 1.0 & S19  \\
        RXC J1314.4-2515$_W$ & 0.247 & 19 & 25 & -11 & 17 & $25\arcsec\times25\arcsec$ & 98 & 0.5 & S19 \\
        \hline
        SPT-CL J2032-5627$_{NW}$ & 0.284 & -- & -- & -- & -- & -- & -- & 1.2 & \\ 
        SPT-CL J2032-5627$_{SE}$ & 0.284 & -- & -- & -- & -- & -- & -- & 0.4 & \\ 
        \hline
        ZwCl 0008.8+5215$_E$ & 0.104 & -- & 25 & -- & -- & $23.5\arcsec\times17.0\arcsec$ & 45 & 0.9 & vW11 \\
        ZwCl 0008.8+5215$_W$ & 0.104 & -- & 10 & -- & -- & $23.5\arcsec\times17.0\arcsec$ & 45 & 0.7 & vW11 \\
        \hline
        ZwCl 1447.2+2619$_N$ & 0.372 & -- & -- & -- & -- & -- & -- & 0.5 & \\
        ZwCl 1447.2+2619$_S$ & 0.372 & -- & -- & -- & -- & -- & -- & 0.8 & \\
        \hline
        ZwCl 2341.1+0000$_N$ & 0.270 & $<5$ & -- & -- & -- & $28\arcsec\times28\arcsec$ & 117 & 0.8 & tw \\
        ZwCl 2341.1+0000$_S$ & 0.270 & $13\pm2$ & 33 & -3 & 35 & $28\arcsec\times28\arcsec$ & 117 & 0.9 & tw \\
		\hline
	\end{tabular}
	}
	\raggedright
		$^a$ We listed only the values obtained for source F in the eastern relic of 8C0212. \\
		$^b$ The only available polarization and Faraday rotation study of this cluster was performed at 2.1 GHz. \\
		$^c$ This value is computed as the median RM value subtracted by the median GRM estimated by \citet{DiGennaro21}, although the authors noticed that this cluster is in a region heavily affected by the Galactic foreground and that the the GRM could be underestimated.  \\
		{\bf References legend}: tw this work, Bo09 \citet{Bonafede09a}, S21 \citet{Stuardi21}, Ka12 \citet{Kale12}, dG22 \citet{deGasperin22}, Li14 \citet{Lindner14}, DG21 \citet{DiGennaro21}, Bo12 \citet{Bonafede12}, J21 \citet{Jones21}, dG15 \citet{deGasperin15}, S19 \citet{Stuardi19}, vW11 \citet{vanWeeren11a}.
\end{table*}

\begin{table*}[!h]
	\caption{Spectral index and Mach number estimates for our compilation of double radio relic galaxy clusters. Column 1: name of the cluster. A subscript letter specifies the relic; Column 2: injection spectral index. Column 3: radio injection Mach number retrieved from the literature (Eq.~\ref{eq:mach_inj}).; Column 4: integrated spectral index; Column 5: radio Mach number derived from the integrated spectral index using Eq.~\ref{eq:mach_int}. We used the $**$ symbol for $\alpha_{\rm int} \leq 1$ since it is incompatible with DSA predictions. $1\sigma$ uncertainties are derived with standard propagation; Column 6: references for injection spectral index and Mach number and for integrated spectral index, as for the legend. For Abell 521$_{NW}$ and PLCK G200.9-28.2$_{NE}$ we give the reference for the discovery but no spectral index is available.}
	\label{tab:double_all_Mach}
	\begin{tabular}{lccccc}
		\hline
		\hline
		Cluster$_\textrm{relic}$ & $\alpha_{\rm inj}$ & $M_{\rm inj}$ & $\alpha_{\rm int}$ & $M_{\rm int}$ & References \\
        \hline
        8C 0212+703$_E$ (ClG 0217+70) & 0.93$\pm$0.08 & 2.4$\pm$0.2 & 1.09$\pm$0.06 & 5$\pm$1 & H21 \\
        8C 0212+703$_W$ (ClG 0217+70) & 0.72$\pm$0.05 & $3.2^{+0.4}_{-0.3}$ & 1.01$\pm$0.05 & ** & H21 \\
        \hline
        Abell 1240$_N$ & 0.94$\pm$0.06 & 2.4$\pm$0.1 & 1.08$\pm$0.05 & 5$\pm$1 & H18  \\
        Abell 1240$_S$ & 0.97$\pm$0.05 & 2.3$\pm$0.1 & 1.13$\pm$0.05  & 4.0$\pm$0.7 & H18 \\
        \hline
        Abell 2146$_N$ & 1.06$\pm$0.09 & 2.1$\pm$0.1 & 1.14$\pm$0.08 & 4$\pm$1 &  H19  \\
        Abell 2146$_S$ & 1.13$\pm$0.06 & 2.0$\pm$0.1 & 1.25$\pm$0.07 & 3.0$\pm$0.4 & H19 \\
        \hline
        Abell 2345$_E$ & -- & -- & 1.29$\pm$0.07 & 2.8$\pm$0.3 & Ge17 \\
        Abell 2345$_W$ & -- & -- & 1.52$\pm$0.08 & 2.2$\pm$0.1 & Ge17 \\
        \hline
        Abell 3186$_{NW}$ (MCXC J0352.4-7401) & -- & -- & 1$\pm$0.1 & ** & D21c  \\
        Abell 3186$_{SE}$ (MCXC J0352.4-7401) & -- & -- & 0.9$\pm$0.1 & ** & D21c \\ 
        \hline
        Abell 3365$_E$ & -- & -- & 0.85$\pm$0.03 & ** & D21c  \\
        Abell 3365$_W$ & -- & -- & 0.76$\pm$0.08 & ** & D21c \\
        \hline
        Abell 3376$_E$ & 0.70$\pm$0.15 & 3.3$\pm$0.3 & 1.33$\pm$0.08 & 2.7$\pm$0.3 & Ka12, C22  \\
        Abell 3376$_W$ & 1.0$\pm$0.2 & 2.2$\pm$0.4 & 1.22$\pm$0.05 & 3.2$\pm$0.3 & Ka12, C22 \\ 
        \hline
        Abell 3667$_{NW}$ & 1.0$\pm$0.1 & 2.2$^{+0.2}_{-0.1}$ & 1.13$\pm$0.02 & 4.0$\pm$0.3 & dG22  \\
        Abell 3667$_{SE}$ & -- & -- & 0.93$\pm$0.03 & ** & dG22 \\
        \hline
        Abell 521$_{NW}$ & -- & -- & -- & -- & Kn22   \\
        Abell 521$_{SE}$ & -- & -- & 1.48$\pm$0.01 & 2.27$\pm$0.02 & Gi08  \\ 
        \hline
        ACT-CL J0102-4915$_{NW}$ (El Gordo) & 0.86$\pm$0.15 & $2.5^{+0.7}_{-0.3}$ & 1.25$\pm$0.04 & 3.0$\pm$0.2 & Li14, tw \\
        ACT-CL J0102-4915$_{E}$ (El Gordo) & -- & -- & 1.06$\pm$0.04 & 6$\pm$2 & Li14, tw \\
        \hline
        CIZA J2242.8+5301$_N$ (Sausage) & 0.86$\pm$0.05 & 2.6$\pm$0.2 & 1.12$\pm$0.03 & 4.2$\pm$0.5 & DG18, Lo20   \\
        CIZA J2242.8+5301$_S$ (Sausage) & 1.09$\pm$0.05 & 2.10$\pm$0.08 & 1.12$\pm$0.07 & 4$\pm$1 & DG18 \\ 
        \hline
        MACS J1752.0+4440$_{NE}$ & 0.6 & 4.6 & 1.16$\pm$0.03 & 3.7$\pm$0.3 & Bo12, vW12 \\
        MACS J1752.0+4440$_{SW}$ & 0.8 & 2.8 & 1.10$\pm$0.05 & 5$\pm$1 & Bo12, vW12 \\
        \hline
        PLCK G200.9-28.2$_{NE}$ & -- & -- & -- & -- & Kn22 \\
        PLCK G200.9-28.2$_{SW}$ & 0.7$\pm$0.2 & 3$\pm$2 & 1.21$\pm$0.15 & 3$\pm$1 & Ka17 \\
        \hline
        PLCK G287.0+32.9$_N$ (PSZ2 G286.98+32.90) & -- & -- & 1.19$\pm$0.03 & 3.4$\pm$0.2 & Ge17 \\
        PLCK G287.0+32.9$_S$ (PSZ2 G286.98+32.90) & -- & -- & 1.36$\pm$0.04 & 2.6$\pm$0.1 & Ge17 \\
        \hline
        PSZ1 G096.89+24.17$_N$ (ZwCl 1856.8+6616) & 0.87$\pm$0.07 & 2.5$\pm$0.2 & 0.92$\pm$0.04 & ** & J21, tw \\
        PSZ1 G096.89+24.17$_S$ (ZwCl 1856.8+6616) & 0.97$\pm$0.07 & 2.3$\pm$0.2 & 1.14$\pm$0.03 & 3.6$\pm$0.4 & J21, tw \\
        \hline
        PSZ1 G108.18-11.53$_{NE}$ & $1.02^{+0.04}_{-0.08}$ & $2.20^{+0.07}_{-0.14}$ & 1.25$\pm$0.02 & 3.0$\pm$0.1 & dG15 \\
        PSZ1 G108.18-11.53$_{SW}$ & $0.952^{+0.09}_{-0.12}$ & $2.3^{+0.2}_{-0.3}$ & 1.28$\pm$0.02 & 2.85$\pm$0.09 & dG15 \\
        \hline
        PSZ2 G233.68+36.14$_N$ & 1.07$\pm$0.11 & $2.1^{+0.2}_{-0.1}$ & 1.31$\pm$0.12 & 2.7$\pm$0.5 & Gh21 \\
        PSZ2 G233.68+36.14$_{SE}$ & 0.67$\pm$0.11 & $3.5^{+2.3}_{-0.7}$ & 0.97$\pm$0.13 & ** & Gh21 \\
        \hline
        RXC J1314.4-2515$_E$ & -- & -- & 1.2$\pm$0.2 & 3$\pm$1 & S19  \\
        RXC J1314.4-2515$_W$ & -- & -- & 1.5$\pm$0.1 & 2.2$\pm$0.2 & S19 \\
        \hline
        SPT-CL J2032-5627$_{NW}$ & -- & -- & 1.2$\pm$0.1 & $3.3\pm0.7$ & D21a \\ 
        SPT-CL J2032-5627$_{SE}$ & -- & -- & 1.5$\pm$0.1 & $2.2\pm0.2$ & D21a \\ 
        \hline
        ZwCl 0008.8+5215$_E$ & 1.2$\pm$0.2 & $2.2^{+0.2}_{-0.1}$ & 1.59$\pm$0.06 & 2.09$\pm$0.08 & vW11 \\
        ZwCl 0008.8+5215$_W$ & 1.0$\pm$0.15 & $2.4^{+0.4}_{-0.2}$ & 1.49$\pm$0.12 & 2.2$\pm$0.2 & vW11 \\
        \hline
        ZwCl 1447.2+2619$_N$ & -- & -- & 1.27$\pm$0.31 & $3\pm1$ & Le22 \\
        ZwCl 1447.2+2619$_S$ & -- & -- & 1.68$\pm$0.3 & $2.0\pm0.3$ & Le22 \\
        \hline
        ZwCl 2341.1+0000$_N$ & -- & $2.4\pm0.4$ & 1.02$\pm$0.02 & ** & Z21, tw \\
        ZwCl 2341.1+0000$_S$ & 1.00$\pm$0.06 & $2.2\pm0.1$ & 0.98$\pm$0.02 & ** & Z21, tw \\
		\hline
	\end{tabular}
	\raggedright
	    \\
		{\bf References legend}: tw this work, H21 \citet{Hoang21}, H18 \citet{Hoang18}, H19 \citet{Hoang19}, Ge17 \citet{George17}, D21c \citet{Duchesne21c}, Ka12 \citet{Kale12}, C22 \citet{Chibueze22}, dG22 \citet{deGasperin22}, Kn22 \citet{Knowles22}, Gi08 \citet{Giacintucci08}, Li14 \citet{Lindner14}, DG18 \citet{Digennaro18}, Lo20 \citet{Loi20}, Bo12 \citet{Bonafede12}, vW12 \citep{vanWeeren12c}, Ka17 \citet{Kale17}, J21 \citet{Jones21}, dG15 \citet{deGasperin15}, Gh21 \citet{Ghirardini21}, S19 \citet{Stuardi19}, D21a \citet{Duchesne21a}, vW11 \citet{vanWeeren11a}, Le22 \citet{Lee22}, Z21 \citet{Zhang21}.
\end{table*}

\subsection{Fractional polarization and depolarization effects}
\label{sec:depol}

High levels of fractional polarization are expected from radio relics, in particular for those seen edge-on, such as double radio relics. This is due to the plane-of-the-sky magnetic field compression operated by the passing shock wave \citep{Ensslin98,Iapichino12}. Recent and advanced MHD simulations show that magnetic field alignment, and therefore high levels of fractional polarization, can be produced by the compression of a randomly oriented magnetic field, which is the natural outcome of the turbulent evolution in the ICM \citep{Wittor19,Dominguez21b}.

Using the simple analytical formula derived by \citet[][Eq. 22]{Ensslin98} and assuming DSA, a radio relic generated by a planar shock wave with Mach number 3 propagating along the plane of the sky reaches an average polarization fraction of 62$\%$. Lower viewing angles lead to lower polarization fractions. Clearly, not all double radio relics are perfectly seen edge-on and projection effects could have a role in determining their polarization fraction. Sometimes this is also suggested by their asymmetry with respect to the merger axis or non-arc-like shape, such as the western relic in A3365 (Fig.~\ref{fig:A3365_nice}) and the relics in ZwCl2341 (Fig.~\ref{fig:ZwCl2341_nice}). However, on average, double radio relics should constitute a more uniform sample with respect to the viewing angle and merger geometry compared to single or multiple radio relic systems \citep{Golovich19b}.

The average fractional polarization of known double radio relics is in the range $9-33 \ \%$ (with average $19 \ \%$)  while the maximum value spans between $10 \ \%$ and $70 \ \%$ (with average $40 \ \%$). Hence, the majority of double radio relics shows lower values of maximum and average polarization with respect to the average value expected for radio relics seen edge-on, if depolarization is not present.

However, both filamentary polarized structures and fractional polarization gradients are found in radio relics when they are observed at high-resolution \citep[i.e. $\leq20$ kpc,][]{DiGennaro21, Rajpurohit22b, deGasperin22}. In particular, \citet{DiGennaro21} found a clear gradient across the Sausage relic with intrinsic fractional polarization decreasing towards the cluster center. This trend can be reproduced by simulations considering a $M = 3$ shock wave propagating through a medium perturbed by decaying subsonic turbulence in the ICM \citep{Dominguez21b}. This trend can be also reproduced with semi-analytical models based on shock compression of a small-scale tangled magnetic field, although this result was found to strongly depend on the magnetic field strength \citep{Hoeft22}. Furthermore, the morphology of simulated polarized emission resembles the structures of the underlying turbulent ICM, creating threads and filaments \citep{Dominguez21b}. The mixing of different polarized structures within the observing beam leads to lower value of average fractional polarization than what is predicted by the analytical formula provided by \citet{Ensslin98}. This effect is particularly significant if the physical scale corresponding to the beam size is of the same order of the reversal (or correlation) scale of magnetic field structures.

Furthermore, linearly polarized emission can be depolarized by several effects, in particular:
\begin{itemize}
\item  beam depolarization, caused by the mixing of several lines-of-sight having different polarization angles within the resolution beam; 
\item  differential Faraday rotation, when a region of space contains relativistic electrons, thermal electrons and regular magnetic fields and the polarization angle of the radiation emitted from the farthest layer is more Faraday-rotated than that from the nearest one;
\item internal and/or external Faraday dispersion, caused by the presence of turbulent and filamentary magnetic fields inside or in front of the radio relic emission that produce an RM dispersion along the line-of-sight and/or within the resolution beam.
\end{itemize}

In principle, the RM-synthesis can overcome both differential Faraday rotation and internal Faraday dispersion because it distinguishes polarized emissions having different Faraday depth along the line-of-sight. In none of our radio relics we detected multiple Faraday depth peaks and therefore we can exclude the presence of multiple emission layers with RM dispersion larger than our resolution in Faraday space, i.e. $45\radm$. However, a smaller Faraday dispersion would be undetectable by our observations and only with a larger bandwidth we would be able to recover it. 

Beam depolarization due to the mixing of intrinsically different polarization angles or to external Faraday dispersion can be avoided only with observations performed at higher resolution. However, high-resolution observations are less sensitive to the faint extended emission that characterizes radio relics. A trade-off between depolarization and loss of sensitivity to the extended emission has to be found in order to observe the polarized emission of radio relics.

\begin{figure}
	\includegraphics[width=1.1\linewidth]{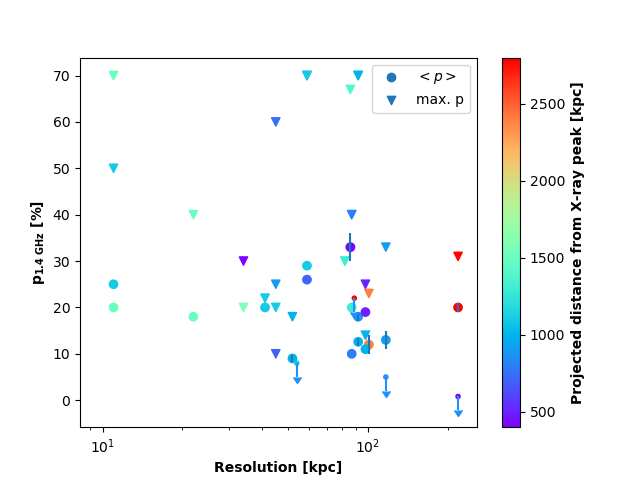}
    \caption{Fractional polarization versus physical resolution of the observation for double radio relics. Each marker represents a single relic. Circles are average fractional polarization computed integrating over the polarized regions of the relic, downwards triangles are the maximum values (therefore one relic can have both measurements in the plot). Arrows are upper limits computed for relics where we did not detect polarization. The color-scale represents the projected distance of the relic from the X-ray peak of the hosting galaxy cluster.}
    \label{fig:pol_res}
\end{figure}

Since beam depolarization should be reduced by high-resolution observations, we want to explore if lower fractional polarization values are found in double relics observed at lower resolution. We listed in Tab.~\ref{tab:double_all} the major and minor axis of the observing beam for each observation and the corresponding physical size. The distribution of fractional polarization values is plotted against the physical resolution of the observation in Fig.~\ref{fig:pol_res}. The markers are color-coded with the distance of each relic from the X-ray centroid of the cluster (also listed in Tab.~\ref{tab:double_all}). We do not observe a strong correlation between the two quantities with a Spearman correlation coefficient of -0.35 using the average polarization values, and 0.02 using the maximum fractional polarization. This suggests that the physical resolution of the observation is not the main driver of depolarization effects in current observations. We do not observe a correlation either with the relic's distance from the cluster center, which should account for larger RM variations within the beam.

\citet{Dominguez21b} showed that considering a subsonic turbulence with power peaking at 50 or 130 kpc, beam depolarization is strong up to a physical resolution of 10-20 kpc (with the average polarization decreasing from $35-65 \ \%$ to $10-40 \ \%$ at 1.5 GHz), while the average fractional polarization remains almost constant at larger resolution beams. Within our double radio relic sample only Abell 3667 has observations that resolve the 20 kpc scale \citep{deGasperin22}. We thus confirm that only a moderate decreasing trend of polarization fraction with resolution is observed for scales larger than $\sim30$ kpc. This is consistent with the simulations of \citet{Dominguez21b}, who considered a turbulent ICM with magnetic field strength of $\sim1 \mu$G within the relics.

The depolarization effect of external Faraday dispersion depends on the observing wavelength and on the RM dispersion experienced by the polarized emission \citep{Burn66}:

\begin{equation}
    p(\lambda) = p_0 e^{-2\sigma_\textrm{RM}^2\lambda^4} \ ,
    \label{eq:externaldep}
\end{equation}

where $p_0$ is the intrinsic polarization fraction at zero wavelength. We can use the RM dispersion computed for the radio relics in this work to verify if external Faraday dispersion is able to account for the observed values of fractional polarization.

In the case of the western relic of 8C0212 we measured $\sigma_\textrm{RM} = 10\radm$, i.e. $14\radm$ in its rest-frame. Considering an intrinsic $p_0 = 62 \ \%$ \citep[computed with $M=3.2$ from][]{Hoang21} we would expect $33 \ \%$ at 1.5 GHz using Eq.~\ref{eq:externaldep} for a perfectly edge-on relic, while we measured a maximum fractional polarization of $23 \ \%$. However, this relic is not very bright and we detected polarization only from the brightest regions while in others the upper limit on the fractional polarization reaches the $30 \ \%$. Therefore, our measurements are in agreement with external depolarization. We do not have information about the ICM distribution at the position of the relics in 8C0212. New X-ray and optical observation are necessary in order to understand the environment of this radio relic, the dynamic that led to its formation and possible projection effects that could further contribute to its depolarization.

The same calculation can be repeated for the eastern radio relic of A3365. We measured $\sigma_\textrm{RM} = 11\radm$ ($13\radm$ in the source rest-frame) and a maximum $p$ of $18 \ \%$. Upper limits in non-detected regions also do not exceed the $20 \ \%$ level. However, in this case an underlying shock wave with Mach number 3.5 was detected with X-ray observations \citet{Urdampilleta21}. Thus, projection effects cannot be the origin of such low fractional polarization. One possibility is that we are detecting polarization only from the external layer of the radio relic, while the RM dispersion is much larger within it. We therefore suggest that internal Faraday dispersion is present with $\sigma_\textrm{RM} < 45\radm$, i.e. with an RM dispersion lower than our resolution in Faraday space. Using the formula for internal Faraday dispersion provided by \citet{Arshakian11}:

\begin{equation}
    p(\lambda) = p_0 \frac{1 - e^{-2\sigma_\textrm{RM}^2\lambda^4}}{2\sigma_\textrm{RM}^2\lambda^4} \ ,
    \label{eq:internaldep}
\end{equation}

we obtain that $\sigma_\textrm{RM} \sim 30\radm$ is sufficient to reduce the fractional polarization to the observed value. This value is consistent with internal RM dispersion found in MHD simulations of radio relics \citep[][see also Sec.~\ref{sec:RMprop}]{Dominguez21b}.

The Faraday depolarization, internal and/or external, could also explain the non-detection in polarization of the bright northern relic in the PLCK287 galaxy cluster. An external $\sigma_\textrm{RM} \geq 30\radm$ or an internal $\sigma_\textrm{RM} \geq 155\radm$ or a combination of the two are able to completely depolarize the signal below the $0.8 \ \%$ level at 1.5 GHz. In this case, the position of the radio relic nearby the galaxy cluster center can clearly account for such level of RM dispersion.

An RM dispersion of $\sigma_\textrm{RM}=22\radm$ ($35\radm$ at the cluster's redshift) as we found for the southern relic of ZwCl2341 is able to explain the low fractional polarization observed for this relic. Moreover, projection effects are likely to lower the polarization fraction of the relics in this system, as also suggested by their non arc-like shape and from the complex X-ray structure of the system \citep{Zhang21}.

In conclusion, at the physical resolution of current observations (i.e., $>30$ kpc) the fractional polarization level of radio relics is already heavily dominated by beam-independent depolarization effects at 1.4 GHz. Both external and internal Faraday depolarization contribute to their observed polarization fraction. This is consistent with simulations in which a tangled magnetic field of strength $\sim1 \mu$G is compressed by a shock wave propagating in a turbulent ICM \citep{Dominguez21b}.

\subsection{Fractional polarization and Mach number}
\label{sec:Mach}

The fractional polarization of radio relics is expected to increase with higher shock Mach numbers. Using again the analytical expression derived by \citet{Ensslin98} for an edge-on relic, the average fractional polarization is expected to increase from $41 \ \%$ up to $62 \ \%$ going from $M$ = 1.5 to $M$ = 3. A much smoother increase is expected for Mach numbers higher than 3, with $<p> = 64 \ \%$ for $M$ = 4.5. A similar trend is obtained by more complex semi-analytical models which also considered the dependence on magnetic field strength \citep{Hoeft22}.

The shock Mach number can be derived from the spectral index of the radio emission assuming particle acceleration via DSA \citep[e.g.,][]{Colafrancesco17}, or from the surface brightness and temperature jump measured from X-ray images \citep[e.g.,][]{Akamatsu13a}. The long-standing problem of the discrepancy between radio- and X-ray-derived Mach numbers (with the radio Mach number being generally higher), recently found a plausible explanation discussed in \citet{Wittor21}. Based on the numerical view of simulated shocks, it is reasonable that rather than a single uniform Mach number, radio relics are characterized by a Mach number distribution which depends on the initial strength of the shock wave and on the turbulent fluctuations in the upstream ICM \citep{Skillman13,Dominguez21a}. While radio observations are more sensitive to the highest Mach numbers within the distribution, which produce the highest radio emissivity, X-ray measurements reflect the average of the distribution \citep[e.g.,][]{Hong15,Wittor21}. Furthermore, X-ray measurements are heavily affected by radio relic's orientation and are biased towards lower values in the case of inclination with respect to the line of sight. For this reason, \citet{Wittor21} concluded that Mach numbers derived from the integrated spectral indexes are more robust.

However in this case, beside the DSA mechanism, a further assumption of planar and stationary shock condition has to be made in order to derive $M$. Under these assumptions the integrated spectral index, $\alpha_\textrm{int}$, is related to the injection spectral index, $\alpha_\textrm{inj}$, by the simple relation

\begin{equation}
    \alpha_\textrm{int} = \alpha_\textrm{inj}+0.5 \ ,
\end{equation}

and the Mach number is:

\begin{equation}
    \label{eq:mach_int}
    M_\textrm{int} = \sqrt{\frac{\alpha_\textrm{int}+1}{\alpha_\textrm{int}-1}} \ ,
\end{equation}

where $\alpha_\textrm{int}>1$.
Alternatively, the radio Mach number can be derived from the injection spectral index measured at the shock front from spatially resolved spectral index maps. In this case 

\begin{equation}
    \label{eq:mach_inj}
    M_\textrm{inj} = \sqrt{\frac{2\alpha_\textrm{inj}+3}{2\alpha_\textrm{inj}-1}} \ .
\end{equation}

While Mach numbers derived from the integrated spectral index can be biased due to simplified assumptions \citep[e.g.,][]{Kang15a} and inaccurate source-subtraction, the estimates derived from the injection spectral index can be biased by coarse spatial resolution, projection effects and misalignment between the radio images.

For the 22 double relic systems, in Tab.~\ref{tab:double_all_Mach} we listed both the Mach number derived from the injection spectral index measured from resolved spectral index maps found in the literature, and the Mach number that we computed with Eq.~\ref{eq:mach_int} using the integrated spectral index values (also listed in the table). In the latter case, we notice that the simple DSA with stationary assumption cannot be applied to 10 radio relics within our sample since they have $\alpha_\textrm{int}\leq1$. This disagreement with the DSA theory may be partially ascribed to the underestimation of the uncertainties on spectral index estimates which, as discussed earlier, can be biased for a number of reasons.

For some of the relics, we computed the integrated spectral index using archival and/or proprietary broad-band observations in order to reduce the uncertainty on the derived Mach number. Flux densities and updated spectral index estimates, together with the plots of the spectra, are reported in the Appendix~\ref{app:A}.

We also notice that often $M_\textrm{inj}<M_\textrm{int}$, with the majority of $M_\textrm{inj}$ being lower than 3. While the injection Mach number show significant variation across the relic, the integrated Mach number is based on the emission-weighted spectral index where higher Mach numbers have more weight. Therefore, the average injection Mach number is often lower than the integrated one. However, with accurate and highly-resolved spectral index maps it is possible to consistently recover the injection and the integrated Mach numbers \citep[see e.g.,][]{Rajpurohit18}.

\begin{figure}
	\includegraphics[width=1.05\linewidth]{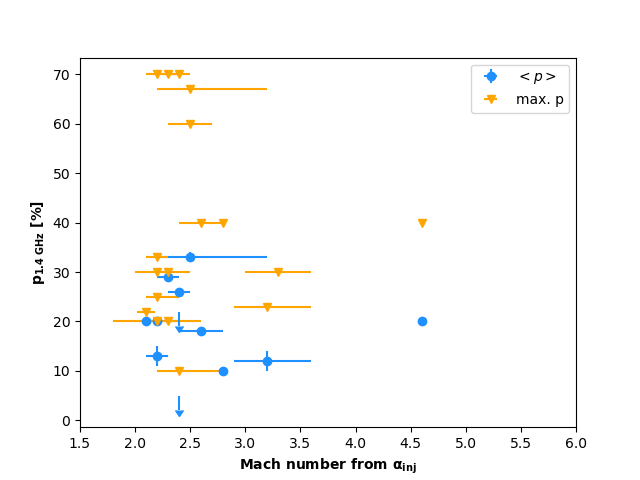}
	\includegraphics[width=1.05\linewidth]{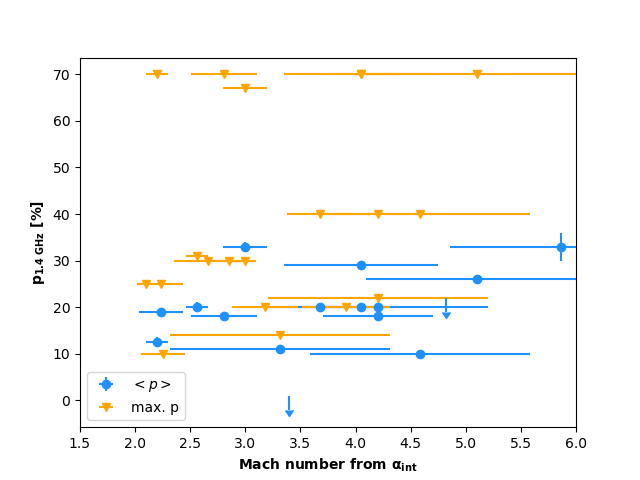}
    \caption{Fractional polarization versus Mach number obtained from the injection spectral index (upper panel) and from the integrated spectral index (bottom panel). Each marker represents a single relic. Blue circles are average fractional polarization computed integrating over the polarized regions of the relic, orange downwards triangles are the maximum values (therefore one relic can have both measurements in the plot). Arrows are upper limits computed for relics where we did not detect polarization. }
    \label{fig:pol_M}
\end{figure}

We searched for a correlation between fractional polarization (or maximum factional polarization value) and Mach number. Plots are shown in Fig.~\ref{fig:pol_M}. We found very weak correlations both between fractional polarization and the Mach number estimated from the injection index (Spearman correlation coefficients -0.23 for $<p>$ and 0.22 for the maximum $p$) and with the Mach number estimated from the integrated radio spectrum (Spearman correlation coefficients 0.29 and 0.24 for $<p>$ the maximum $p$, respectively). The weak correlation can be partially due to the large uncertainties on Mach number estimates and on the aforementioned possible bias present in both methods. However, the Spearman correlation coefficient is positive for both $<p>$ and the maximum $p$ only for the integrated Mach number, and it reaches the highest value in the correlation between $M_{\rm int}$ and $<p>$. We interpret this fact as a suggestion that the Mach number estimated from the integrated spectral index is in fact more robust, as suggested by \citet{Wittor19}.

Furthermore, while the majority of injection-derived Mach numbers are lower than 3, the distribution of $M_\textrm{int}$ is shifted towards higher values where we expect a weaker correlation between fractional polarization and Mach number \citep[see Fig. 1 in][]{Hoeft22}. If the Mach number distribution would be the one described by $M_\textrm{inj}$ we would have found a stronger correlation. This suggests that the bulk of of double radio relics reaches maximum Mach numbers $M>2.5-3$, as required by particle acceleration models from the thermal pool which would require an unrealistically high acceleration efficiency for $M\leq2$ \citep{Botteon20a,Dominguez21a}. Also the fractional polarization observed in radio relics would be difficult to reproduce with $M\leq2$ due to the low level of magnetic field compression \citep{Dominguez21b}.

Overall, the observed weak positive trend of fractional polarization with Mach number can be explained in the context of turbulent magnetic field compression for $M>2.5-3$. The magnetic field strength has also a major role in determining the fractional polarization level of radio relics due to its impact on particle aging \citep{Hoeft22}.

\subsection{Faraday rotation properties}
\label{sec:RMprop}

The average (or median) residual RM of double radio relics in the cluster's rest-frame spans between $-13 \radm$ and $38 \radm$, with the only exception of CIZA J2242.8+5301 ($-86 \radm$) for which an high residual contribution from the Galactic RM is very likely \citep{DiGennaro21}. The measured RRM are in the lower range of relics RMs predicted by MHD cosmological simulations which only take into account the external ICM contribution and spans in the range 10-100$\radm$ \citep{Wittor19}. This is consistent with double radio relics being seen edge-on and lying in the outskirts of galaxy clusters, therefore crossing a small Faraday-rotating volume.

The rest-frame RM dispersion spans between $8 \radm$ and $35 \radm$, perfectly consistent with MHD cosmological simulations which predict $\sigma_\textrm{RM}$ of few tens for edge-on relics and $\sigma_\textrm{RM}$ of few hundreds for face-one relics \citep{Wittor19}. However, due to external Faraday dispersion, $\sigma_\textrm{RM}$ larger than $40 \radm$ would be undetectable at 1.5 GHz, because the signal would be totally depolarized. Higher-frequency observations are needed to exclude a possible observing bias.

\citet{Dominguez21b} simulated the internal RM of radio relic within a (200 kpc)$^3$ volume. They found that the RM dispersion within relics depends on the pre-shock turbulent conditions of the ICM. In particular, they found that a subsonic turbulence with power peaking at larger scales produces a higher internal RM dispersion as a consequence of a broader magnetic field distribution. They also found that the internal RM distribution tends to narrow when taking into account only brighter radio emitting regions, as expected for polarization measurements. 

We did not detect internal Faraday rotation with RM dispersion larger than $45 \radm$ in our four galaxy clusters. We infer the presence of internal Faraday rotation with $\sigma_\textrm{RM} \sim 30 \radm$ in the eastern relic of A3365 due to its strong depolarization. The only double radio relic with internal Faraday rotation detected in the literature is the western radio relic of RXC J1314.4-2515 which shows an internal RM dispersion of $\sim100 \radm$ \citep{Stuardi19}. An internal RM dispersion lower than $100 \radm$ is line with simulations which account for a pre-shock turbulence with power peaking at $\sim 50$ kpc while larger scale turbulence would imply larger internal RM dispersion \citep[see Fig.16][]{Dominguez21b}. In these simulations the magnetic field strength is 1.5 $\mu$G. 

\begin{figure}
	\includegraphics[width=1.1\linewidth]{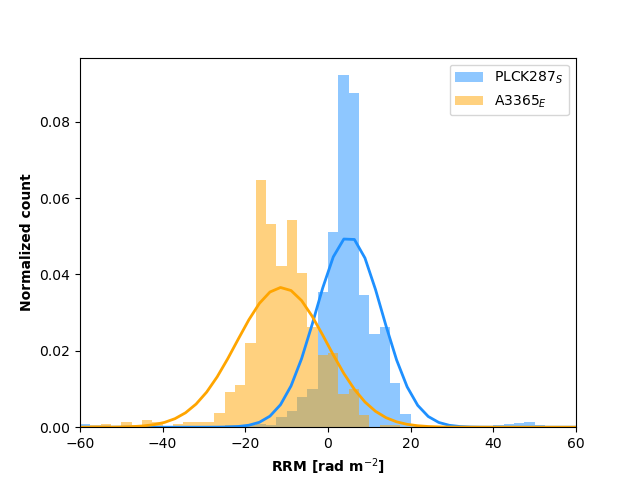}
    \caption{Normalized Residual Rotation Measure distributions (in the observer rest-frame) of the eastern relic of A3365 (orange) and of the southern relic of PLCK287 (blue). The two distribution are compared with Gaussian distributions having the same mean and standard deviation (blue and orange lines) }
    \label{fig:hist_RM}
\end{figure}

Both \citet{Wittor19} and \citet{Dominguez21b} found an asymmetric and non-Gaussian RM distribution in radio relics. This was previously noticed by \citet{Vazza18}, who analyzed highly resolved MHD simulations of entire galaxy clusters. They noticed that the non-Gaussian behavior increases with the simulation resolution and with distance from the cluster center. This resulted in higher RMs values with respect to a Gaussian distribution that could affect the magnetic field estimate derived from RM modeling. We show the RM distributions of the southern relic of PLCK287 and the eastern relic of A3365 compared to their fit with a Gaussian distributions in Fig.~\ref{fig:hist_RM}. Both RRM distributions are non-Gaussian, similarly to what is found in other double radio relics \citep{Stuardi19,Stuardi21,DiGennaro21}. The distribution of RMs is also non-symmetric for Abell 3365. Compared to Gaussian distributions, the observed ones are more peaked and skewed to higher RMs. A similar result was also found by \citet{Vazza18}.

\begin{figure}
	\includegraphics[width=1.05\linewidth]{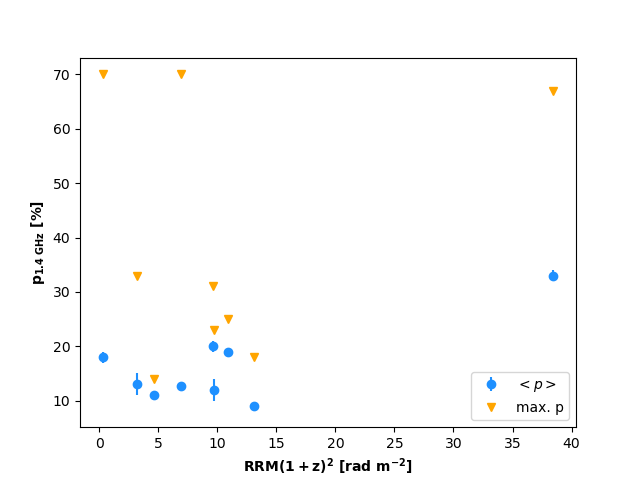}
	\includegraphics[width=1.05\linewidth]{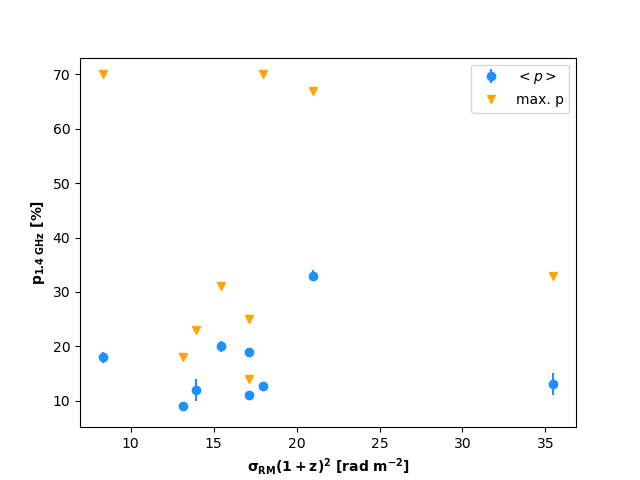}
    \caption{Fractional polarization versus intrinsic residual RM (RRM, i.e. corrected for Galactic contribution and cosmological shift, upper panel) and intrinsic RM dispersion (bottom panel). Blue circles are average fractional polarization computed by integrating over the polarized regions of the relic, orange downwards triangles are the maximum values (therefore one relic can have both measurements in the plot).}
    \label{fig:pol_RM}
\end{figure}

In Fig.~\ref{fig:pol_RM} we show the fractional polarization of double relics at 1.4 GHz as a function of intrinsic RM and $\sigma_{RM}$. We did not find significant correlation between the considered quantities. This confirms that different depolarization effects (differential Faraday rotation, internal and external Faraday dispersion) together contribute to the final fractional polarization observed at 1.4 GHz.

We observed a very good alignment of magnetic field line direction with the main axis of the southern radio relic of PLCK287 on Mpc-scales (see Fig.~\ref{fig:PLCK287_pol}). The alignment is less clear for the eastern relic in A3365 (Fig.~\ref{fig:A3365_pol}). MHD simulations are able to reproduce the magnetic field aliment with the shock front for edge-on relics but only on small scales, i.e. $\leq 200$ kpc \citep{Skillman13,Wittor19}. \citet{Dominguez21b} showed that the magnetic field alignment can increase with decreasing resolution of the radio observation since larger resolution elements weight more the brightest regions where the polarization vector is aligned with the shock normal. We notice that PLCK287 is the cluster for which we have the lowest physical resolution. The fact that many radio relics show ordered magnetic field lines on Mpc-scale agrees with our finding that the physical resolution of current observations (> 30 kpc) is not strongly affecting their fractional polarization properties (Sec.~\ref{sec:depol}). The turbulent scales causing depolarization and magnetic field misalignment are already resolved-out above 20 kpc and we can only observe the brightest regions where magnetic field vectors are aligned. However, simulations show that the level of magnetic field alignment depends also on the physical scale of the upstream turbulence, therefore the differences that we observed between radio relics could also reflect different physical conditions of the ICM.

\section{Conclusions}
\label{sec:conclusion}

In this paper we produced and analyzed polarization and Faraday rotation images of four famous double radio relics galaxy clusters in the 1-2 GHz frequency range using JVLA observations. For our polarization analysis we used RM synthesis \citep{Brentjens05}. With this work we almost doubled the number of double radio relics with available Faraday rotation information. Among our sample, we detected extended polarized emission only from two relics, while the remaining relics are either totally depolarized at 1.5 GHz, or show only few patches of polarized emission. We focused on the possible origin of depolarization. In particular:
\begin{itemize}
        \item \textbf{8C 0212+703 (ClG 0217+70).} In the 8C0212 galaxy cluster we detected a maximum fractional polarization of $23 \ \%$ from the western radio relic (source C) confirming its identification as a radio relic \citep{Hoang21}. We also detected polarization from the lobes of the radio galaxy close to the eastern radio relic (source E1) while the other parts of the relic (sources D, F and G) are undetected in polarization with $28 \ \%$ upper limit to the fractional polarization.
        \item \textbf{Abell 3365.} We detected a low level of fractional polarization (i.e., $<18 \ \%$) and only from the eastern radio relic. Since the external RM dispersion measured is low ($13 \radm$ in the source rest-frame) we suggest that a possible cause of depolarization is internal Faraday dispersion with $\sigma_\textrm{RM} \sim 30 \radm$. Magnetic field vectors are not aligned with the main axis of the relic along the whole relic extent.
        \item \textbf{PLCK G287.0+32.9 (PSZ2 G286.98+32.90).} Only the southern relic of PLCK287 is detected in polarization with $<p> = 20\pm1 \ \%$, and the magnetic field vectors are well aligned with its main axis. Notably, also the faint upstream extension of the relic is polarized. The northern relic is likely depolarized at 1.5 GHz, due its proximity to the cluster center \citep{Bonafede14}. The connection of the norther relic with the large radio galaxy in the north-east is supported by the similarity of its RM properties and the one of the closest cluster galaxy.
        \item \textbf{ZwCl 2341.1+0000.} The northern relic of ZwCl2341 is unpolarized, with fractional polarization upper limit of $5 \%$. This is consistent with previous 2-4 GHz observations \citep{Benson17}. We concluded that previous low-resolution 1.4 GHz observations that found higher fractional polarization values were contaminated by the nearby head-tail radio galaxy \citep{Giovannini10}. The southern relic shows few patches of polarized emission with $<p> = 13\pm2 \ \%$. Projection effects are likely to play an important role in the depolarization of these relics, as suggested by the disturbed X-ray morphology of the cluster \citep{Zhang21}, merger components along the line-of-sight \citep{Golovich19b} and high $\sigma_\textrm{RM}$ value (i.e., $35 \radm$ in the source rest-frame).
\end{itemize}

In order to place our results into a broader context, we made an updated compilation of all double radio relics known to date and we statistically analyzed their polarization and Faraday rotation properties at 1.4 GHz. We listed 22 double radio relics: 15 have polarization information.
These are our main conclusions:
\begin{itemize}
        \item almost all radio double radio relics in the literature have been observed with physical resolution coarser than 30 kpc. In this resolution range, we found a moderate decreasing trend of the average fractional polarization of radio relics with the physical size of the observing beam. This is consistent with simulation suggesting the presence of turbulence with the physical scale peaking between 50 and 130 kpc in the upstream ICM which causes beam-dependent depolarization only up to $\sim20$ kpc \citep{Dominguez21b}.
        \item Both external and internal Faraday dispersion contribute to the observed polarization fraction at 1.4 GHz. In particular, internal Faraday dispersion with $\sigma_\textrm{RM} < 45 \radm$ should be accounted for in order to explain the low fractional polarization of some double radio relics, since the detected external RM dispersion is not enough to depolarize them to the observed value.
        \item We found a weak positive correlation between the fractional polarization of relics and shock Mach number estimated from the integrated spectrum of radio relics (Spearman coefficient $\sim 0.3$). Such a weak correlation is expected for Mach numbers higher than 2.5, while for lower values a stronger correlation would be expected \citep{Hoeft22}. This suggests that most radio relics reach a maximum Mach number higher than 2.5 and that Mach number estimates from X-ray or from injection radio spectral indexes are biased towards lower values. This founding would help explaining the origin or radio relics, since, with the standard DSA from the thermal pool, Mach numbers lower than 2.5 are not expected to generate the radio luminosity and the polarization fraction observed from radio relics \citep{Botteon20a,Dominguez21a,Dominguez21b}. However, we notice that the large uncertainties on Mach number estimates still prevent strong conclusions.
        \item Although the number of radio relics with available Faraday rotation information is still low, we found that the global RM properties of double radio relics are well reproduced by the state-of-the-art MHD simulations \citep{Wittor19}. Both observed RM and $\sigma_\textrm{RM}$ of double radio relics are consistent with what is expected from edge-on relics in cosmological MHD simulations. The amount of internal Faraday rotation observed from double radio relics can be explained by the presence of
        a turbulent ICM up-stream of the shocks with power peaking at $\sim 50$ kpc scales and turbulent magnetic fields of strength $\sim1 \mu$G, as simulated by \citet{Dominguez21b}. We confirm that the RM distribution of radio relics is non-Gaussian and that magnetic field lines appear more aligned at lower resolution.
\end{itemize}

\begin{acknowledgements}
C.S. and A.B. acknowledge support from the MIUR grant FARE ``SMS" and from the ERC-StG DRANOEL, n. 714245. F.V. acknowledges support from the ERC-StG MAGCOW, n. 714196. R.J.vW. acknowledges support from the ERC Starting Grant ClusterWeb n. 804208. We thank the anonymous referee for useful suggestions.
\end{acknowledgements}

%
%

\bibliographystyle{aa} 
\bibliography{my_bib} 

\begin{appendix} 

\section{Broad-band integrated radio spectra}
\label{app:A}

In order to reduce the uncertainties on the Mach number estimates used in Sec.~\ref{sec:Mach}, we required that all radio relics with polarization information have spectral index estimates with uncertainties lower than 0.1. For this reason, we computed the broad-band integrated spectral index of three double radio relics, namely PSZ1 G096.89+24.17 (a.k.a. ZwCl 1856.8+6616), El Gordo, and ZwCl 2341.1+0000. We used both archival and proprietary data from which we derived flux density measurements. The uncertainty on the flux density were computed taking into account both statistical noise and calibration errors:

\begin{equation}
    \sigma_S=\sqrt{(\delta S \times S)^2+(\sigma\times\sqrt{n_{\rm beam}})^2}~,
\end{equation}

where the calibration error, $\delta$, is $5 \ \%$ for JVLA and $10 \ \%$ for GMRT and LOFAR observations, $\sigma$ is the rms noise of the image, and $n_{\rm beam}$ is the number of resolution elements contained in the region used to measure the flux density. We then computed the spectral index and its uncertainty with a standard power-law fitting.

\begin{figure*}[h]
    \centering
	\includegraphics[width=0.45\textwidth]{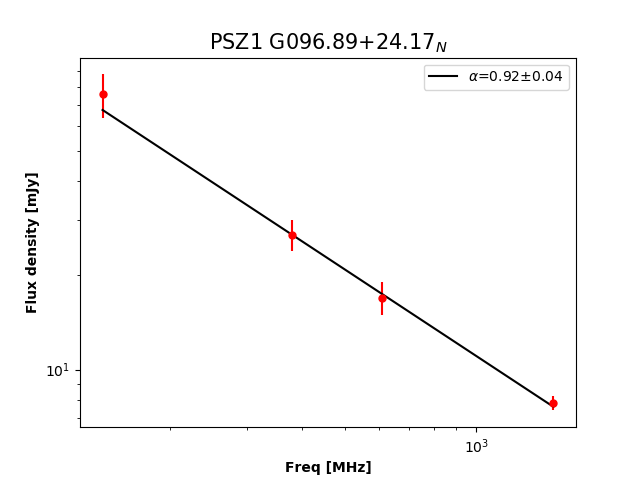}
	\includegraphics[width=0.45\textwidth]{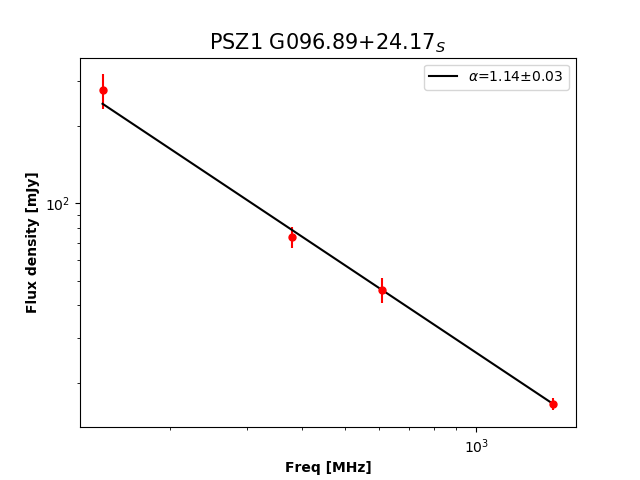}
	\includegraphics[width=0.45\textwidth]{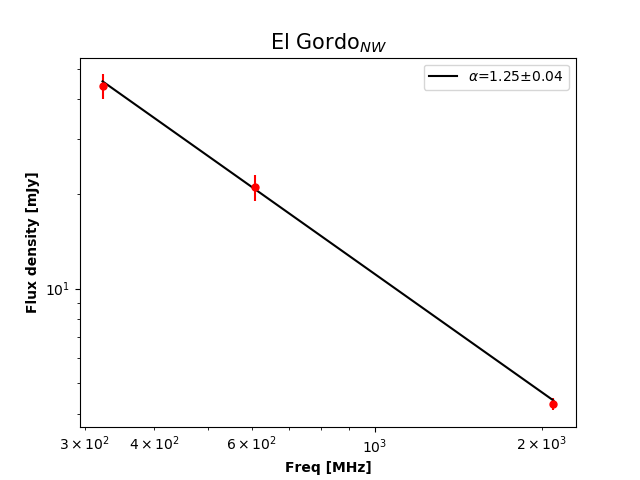}
	\includegraphics[width=0.45\textwidth]{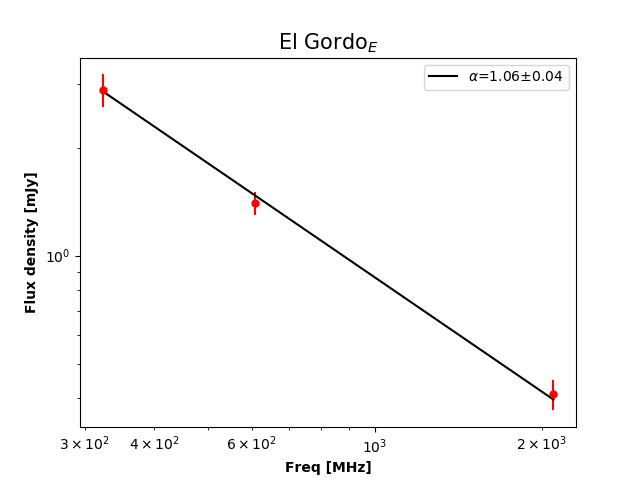}
	\includegraphics[width=0.45\textwidth]{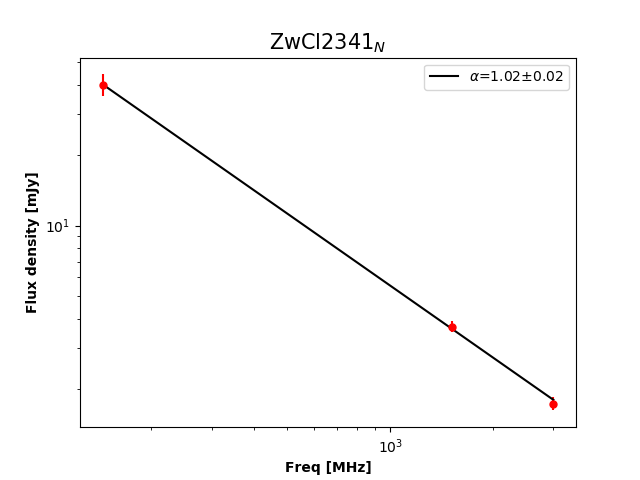}
	\includegraphics[width=0.45\textwidth]{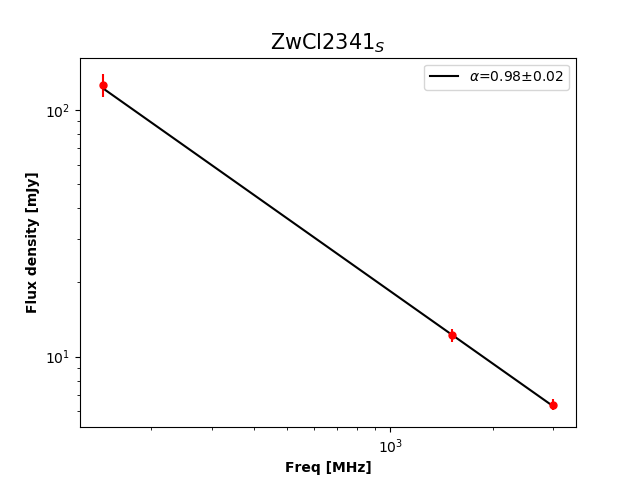}
    \caption{Power-law fit to the flux density measurements listed in Tab.~\ref{tab:flux_PSZ096}, Tab.~\ref{tab:flux_ElGordo} and Tab.~\ref{tab:flux_ZwCl2341}}.
    \label{fig:alpha_fit}
\end{figure*}

For PSZ1 G096.89+24.17, we used flux density measurements at 140 MHz and 1.5 GHz from \citet{Jones21} and we computed 380 and 610 MHz flux densities from proprietary GMRT observations (Rajpurohit et al. in preparation). For El Gordo we measured the flux density at 325 and 610 MHz using archival observations (observed during 2017), and we used the 2.1 GHz measurement from \citet{Lindner14}. For ZwCl2341 we used the 1.5 GHz images published in this work, 3 GHz measurements from \citet{Benson17} and 144 MHz flux densities form proprietary LOFAR data (Hoang et al. in preparation). Flux density measurements and resulting integrated spectral index are listed in Tab.~\ref{tab:flux_PSZ096}, Tab.~\ref{tab:flux_ElGordo}, and Tab.~\ref{tab:flux_ZwCl2341} for PSZ1 G096.89+24.17, El Gordo, and ZwCl 2341.1+0000, respectively. Power-law fits are displayed in Fig.~\ref{fig:alpha_fit}.

We notice that for three clusters (namely, Abell 3365, El Gordo and RXC J1314.4-2515) we attempted a spectral fitting using the enhanced imaging products released by the MeerKAT Galaxy Cluster Legacy Survey \citep{Knowles22}. However, we found that the flux densities measured from these survey were inconsistent with other measurements, leading to unreliable steep spectral indexes ($>2$). Hence, we did not include MeerKAT data here.

\begin{table}
    \centering
	\caption{Broad-band spectrum for PSZ1 G096.89+24.17. Column 1: radio relic; Column 2: frequency; Column 3: flux density with 1$\sigma$ uncertainty; Column 4: spectral index derived from power-law fitting with associated uncertainty.}
	\label{tab:flux_PSZ096}
	\begin{tabular}{lccc}
		\hline
		\hline
		\multicolumn{1}{c}{Relic} & \multicolumn{1}{c}{$\nu$ [MHz]} & \multicolumn{1}{c}{$S_{\nu}$ [mJy]} & \multicolumn{1}{c}{$\alpha$} \\
        \hline
        North & 140 & 76$\pm$12 & 0.92$\pm$0.04\\
         & 380 & 27$\pm3$ & \\
         & 610 & 17$\pm$2 & \\
         & 1500 & 7.8$\pm$0.4 & \\
        South & 140 & 276$\pm$42 & 1.14$\pm$0.03 \\
         & 380 & 74$\pm$7 & \\
         & 610 & 46$\pm$5 & \\
         & 1500 & 16.5$\pm$0.9 & \\
		\hline
	\end{tabular}
\end{table}

\begin{table}
    \centering
	\caption{Broad-band spectrum for El Gordo. Column 1: radio relic; Column 2: frequency; Column 3: flux density with 1$\sigma$ uncertainty; Column 4: spectral index derived from power-law fitting with associated uncertainty.}
	\label{tab:flux_ElGordo}
	\begin{tabular}{lccc}
		\hline
		\hline
		\multicolumn{1}{c}{Relic} & \multicolumn{1}{c}{$\nu$ [MHz]} & \multicolumn{1}{c}{$S_{\nu}$ [mJy]} & \multicolumn{1}{c}{$\alpha$} \\
        \hline
        North West & 325 & 44$\pm$4 & 1.25$\pm$0.04\\
         & 610 & 21$\pm2$ & \\
         & 2100 & 4.3$\pm$0.2 & \\
        East & 325 & 2.9$\pm$0.3 & 1.06$\pm$0.04 \\
         & 610 & 1.4$\pm$0.1 & \\
         & 2100 & 0.41$\pm$0.04 & \\
		\hline
	\end{tabular}
\end{table}

\begin{table}
    \centering
	\caption{Broad-band spectrum for ZwCl 2341.1+0000. Column 1: radio relic; Column 2: frequency; Column 3: flux density with 1$\sigma$ uncertainty; Column 4: spectral index derived from power-law fitting with associated uncertainty.}
	\label{tab:flux_ZwCl2341}
	\begin{tabular}{lccc}
		\hline
		\hline
		\multicolumn{1}{c}{Relic} & \multicolumn{1}{c}{$\nu$ [MHz]} & \multicolumn{1}{c}{$S_{\nu}$ [mJy]} & \multicolumn{1}{c}{$\alpha$} \\
        \hline
        North & 144 & 40$\pm$4 & 1.02$\pm$0.02\\
         & 1500 & 3.7$\pm0.2$ & \\
         & 3000 & 1.7$\pm$0.1 & \\
        South & 144 & 126$\pm$13 & 0.98$\pm$0.02 \\
         & 1500 & 12.2$\pm$0.7 & \\
         & 3000 & 6.4$\pm$0.3 & \\
		\hline
	\end{tabular}
\end{table}

\end{appendix}

\end{document}